\documentclass[useAMS,usenatbib,fleqn]{mn2e}

\usepackage{times}
\usepackage{natbib}
\usepackage[english]{babel}
\usepackage[utf8]{inputenc}
\usepackage{amsmath}
\usepackage{graphicx}
\usepackage[colorinlistoftodos]{todonotes}

\def \mnras {MNRAS}
\def \aj {AJ}
\def \apj {ApJ}
\def \apjs {ApJS}
\def \apjl {ApJ}

\def \pasa {Publ.\ Astron.\ Soc.\ Australia}
\def \pasp {PASP}
\def \aap {A\&A}

\def \apss {Astrophysics and Space Science}

\def \sifa {1}
\def \caastro {2}
\def \aao {3}
\def \anu {4}
\def \unc {{5}}
\def \hawaii {{6}}
\def \ljmu {7}
\def \swinburne {{8}}
\def \queensland {{9}}
\def \uwa {{10}}
\def \standrews {{11}}
\def \durham {{12}}
\def \ipos {{13}}
\def \eso {{14}}
\def \macquariephysics {{15}}
\def \macquariephotonics {{16}}
\def \melbourne {{17}}
\def \aip {{18}}

\title{The SAMI Galaxy Survey: Early Data Release}

\author[J.\,T.~Allen et al.]
{\parbox{\textwidth}{\raggedright J.\,T.~Allen$^{\sifa,\caastro}$\thanks{j.allen@physics.usyd.edu.au},
S.\,M.~Croom$^{\sifa,\caastro}$,
I.\,S.~Konstantopoulos$^{\aao,\caastro}$,
J.\,J.~Bryant$^{\sifa,\caastro,\aao}$,
R.~Sharp$^{\anu}$,
G.\,N.~Cecil$^\unc$,
L.\,M.\,R.~Fogarty$^{\sifa,\caastro}$,
C.~Foster$^\aao$,
A.\,W.~Green$^{\aao}$,
I.-T.~Ho$^\hawaii$,
M.\,S.~Owers$^{\aao}$,
A.\,L.~Schaefer$^{\sifa,\caastro,\aao}$,
N.~Scott$^{\sifa,\caastro}$,
A.\,E.~Bauer$^\aao$,
I.~Baldry$^{\ljmu}$,
L.\,A.~Barnes$^\sifa$,
J.~Bland-Hawthorn$^{\sifa,\caastro}$,
J.\,V.~Bloom$^{\sifa,\caastro}$,
S.~Brough$^\aao$,
M.~Colless$^\anu$,
L.~Cortese$^\swinburne$,
W.\,J.~Couch$^{\aao}$,
M.\,J.~Drinkwater$^\queensland$,
S.\,P.~Driver$^{\uwa,\standrews}$,
M.~Goodwin$^{\aao}$,
M.\,L.\,P.~Gunawardhana$^\durham$,
E.\,J.~Hampton$^\anu$,
A.\,M.~Hopkins$^\aao$,
L.\,J.~Kewley$^\anu$,
J.\,S.~Lawrence$^{\aao}$,
S.\,G.~Leon-Saval$^\ipos$,
J.~Liske$^{\eso}$,
\'A.\,R.~L\'opez-S\'anchez$^{\aao,\macquariephysics}$,
N.\,P.\,F.~Lorente$^\aao$,
R.~McElroy$^{\sifa,\caastro}$,
A.\,M.~Medling$^\anu$,
J.~Mould$^\swinburne$,
P.~Norberg$^{\durham}$,
Q.\,A.~Parker$^{\aao,\macquariephysics,\macquariephotonics}$,
C.~Power$^{\uwa,\caastro}$,
M.\,B.~Pracy$^\sifa$,
S.\,N.~Richards$^{\sifa,\caastro,\aao}$,
A.\,S.\,G.~Robotham$^{\uwa}$,
S.\,M.~Sweet$^{\anu,\queensland}$,
E.\,N.~Taylor$^\melbourne$,
A.\,D.~Thomas$^\queensland$,
C.~Tonini$^\swinburne$,
C.\,J.~Walcher$^\aip$
}\vspace{0.4cm}\\
\parbox{\textwidth}{$^{\sifa}$ Sydney Institute for Astronomy (SIfA), School of Physics, The University of Sydney, NSW 2006, Australia
\\$^{\caastro}$ ARC Centre of Excellence for All-sky Astrophysics (CAASTRO)
\\$^{\aao}$ Australian Astronomical Observatory, PO Box 915, North Ryde, NSW 1670, Australia
\\$^{\anu}$ Research School of Astronomy \& Astrophysics, Australian National University, Mount Stromlo Observatory, Cotter road, Weston Creek, ACT 2611, Australia
\\$^\unc$ Department of Physics and Astronomy, University of North Carolina, Chapel Hill, NC 27599, USA
\\$^\hawaii$ Institute for Astronomy, University of Hawaii, 2680 Woodlawn Drive, Honolulu, HI 96822, USA
\\$^{\ljmu}$ Astrophysics Research Institute, Liverpool John Moores University, IC2, Liverpool Science Park, 146 Brownlow Hill, Liverpool L3~5RF, UK
\\$^\swinburne$ Centre for Astrophysics and Supercomputing, Swinburne University of Technology, Hawthorn, VIC 3122, Australia
\\$^{\queensland}$ School of Mathematics and Physics, University of Queensland, QLD 4072, Australia
\\$^{\uwa}$ International Centre for Radio Astronomy Research, University of Western Australia, 35 Stirling Highway, Crawley, WA 6009, Australia
\\$^{\standrews}$ SUPA, School of Physics and Astronomy, University of St Andrews, North Haugh, KY16 9SS, UK
\\$^{\durham}$ ICC, Department of Physics, Durham University, South Road, Durham DH1 3LE, UK
\\$^\ipos$ Institute of Photonics and Optical Science (IPOS), School of Physics, The University of Sydney, NSW 2006, Australia
\\$^{\eso}$ European Southern Observatory, Karl-Schwarzschild-Str.\ 2, D-85748 Garching, Germany
\\$^\macquariephysics$ Department of Physics and Astronomy, Macquarie University, NSW 2109, Australia
\\$^\macquariephotonics$ Research Centre for Astronomy, Astrophysics and Astrophotonics, Macquarie University, Sydney, NSW 2109 Australia
\\$^\melbourne$ School of Physics, The University of Melbourne, VIC 3010, Australia
\\$^\aip$ Leibniz-Institut f\"ur Astrophysik Potsdam (AIP), An der Sternwarte 16, D-14482 Potsdam, Germany
}}

\date{\today}

\begin{document}
\maketitle


\begin{abstract}
We present the Early Data Release of the Sydney--AAO Multi-object Integral field spectrograph (SAMI) Galaxy Survey. The SAMI Galaxy Survey is an ongoing integral field spectroscopic survey of $\sim$3400 low-redshift ($z<0.12$) galaxies, covering galaxies in the field and in groups within the Galaxy And Mass Assembly (GAMA) survey regions, and a sample of galaxies in clusters.

In the Early Data Release, we publicly release the fully calibrated datacubes for a representative selection of 107 galaxies drawn from the GAMA regions, along with information about these galaxies from the GAMA catalogues. All datacubes for the Early Data Release galaxies can be downloaded individually or as a set from the SAMI Galaxy Survey website.

In this paper we also assess the quality of the pipeline used to reduce the SAMI data, giving metrics that quantify its performance at all stages in processing the raw data into calibrated datacubes. The pipeline gives excellent results throughout, with typical sky subtraction residuals in the continuum of 0.9--1.2 per cent, a relative flux calibration uncertainty of 4.1 per cent (systematic) plus 4.3 per cent (statistical), and atmospheric dispersion removed with an accuracy of 0\farcs09, less than a fifth of a spaxel.
\end{abstract}

\begin{keywords}
galaxies: evolution -- galaxies: kinematics and dynamics -- galaxies: structure -- techniques: imaging spectroscopy
\end{keywords}

\section{Introduction}

Spectroscopic surveys of galaxies -- e.g.\ the 2-degree Field Galaxy Redshift Survey (2dFGRS; \citealt{Colless01}), 6-degree Field Galaxy Survey (6dFGS; \citealt{Jones09}), Sloan Digital Sky Survey (SDSS; \citealt{York00}) and Galaxy And Mass Assembly (GAMA; \citealt{Driver09,Driver11}) survey -- are immensely powerful tools for investigating galaxy formation and evolution. Optical spectroscopy provides measurements of properties such as the star formation rate (SFR), stellar ages, metallicity of gas and stars, and dust extinction. Emission from other processes such as active galactic nuclei (AGN) and shock-heated gas is also measured. Multi-object spectrographs have allowed us to collate these properties for large samples of galaxies and search for correlations that illuminate the underlying physics of galaxy evolution, as well as find outliers that can be used to test models under the most extreme conditions. Surveys of this type have shed new light on the role of environment in star formation (e.g.\ \citealt{Lewis02,Wijesinghe12}), the relationship between supermassive black holes and their host galaxies (e.g.\ \citealt{Shen08}), the mass--metallicity--SFR relation (e.g.\ \citealt{Mannucci10,LaraLopez13}) and many other aspects of extragalactic astrophysics.

Notwithstanding the great achievements of traditional spectroscopic surveys, the use of a single fibre or slit to observe each galaxy limits the available information. A clear example of this limitation is that an SFR measured from a single spectrum must be aperture corrected to recover the global SFR, introducing an extra uncertainty \citep{Hopkins03}. In many cases, the exact location of the fibre or slit within the galaxy is not known, increasing the uncertainty from aperture effects. More critically, single-fibre observations can only provide a single measurement of any property for the entire galaxy, and so are inherently unable to probe spatial variations in SFR, ionisation state, stellar populations or any other parameter. Similarly, they cannot provide any information about the internal kinematics of the galaxy other than an integrated velocity dispersion.

These limitations are addressed by the use of spatially-resolved spectroscopy, based on integral field units (IFUs). Measuring the spatially resolved properties of galaxies has opened up a new direction in the study of galaxy evolution: studies of turbulent star-forming discs \citep{Genzel08}, galactic winds from AGN and star formation \citep{Sharp10b}, the environmental dependence of stellar kinematics \citep{Cappellari11b} and star formation distributions \citep{Brough13}, and age and metallicity gradients in stellar populations \citep{SanchezBlazquez14} provide only a small sample of the contributions of integral field spectroscopy.

Until now, technical limitations have meant that almost all IFU instruments have been monolithic, viewing a single galaxy at a time. As a result, observing large numbers of galaxies has been difficult and time-consuming. The largest samples of IFU observations available to date have been the ATLAS-3D survey, with 260 galaxies \citep{Cappellari11a}, and the ongoing Calar Alto Legacy Integral Field Area (CALIFA) survey, with a target of 600 galaxies \citep{Sanchez12}. Surveys of this size are already breaking new ground in our understanding of galaxy evolution, but with monolithic IFUs they cannot be scaled up much further.

The next step forward for integral field spectroscopic surveys is to use multiplexed IFUs capable of observing multiple galaxies simultaneously. The first such instrument, the Fibre Large Array Multi Element Spectrograph (FLAMES; \citealt{Pasquini02}) on the 8-m Very Large Telescope (VLT) has 15 deployable IFUs with 20 spaxels (spatial elements) each and $2\times3\arcsec$ fields of view. Despite the small number of spaxels in each IFU, FLAMES has demonstrated the power of multiplexed IFU systems, for example showing the increased kinematic complexity of galaxies at medium redshift ($z\sim0.6$; \citealt{Flores06,Yang08}).

For studies at low redshift ($z<\sim0.2$), IFUs with larger fields of view are required to probe the full extent of each galaxy. The Sydney--AAO Multi-object Integral field spectrograph (SAMI; \citealt{Croom12}) on the 3.9-m Anglo-Australian Telescope (AAT) is the first instrument with this capability, having 13 IFUs each with a field of view of $\approx$15\arcsec. SAMI makes use of hexabundles, bundles of 61 fibres that are fused together and have a high optical performance and $\sim$75 per cent filling factor \citep{BlandHawthorn11,Bryant11,Bryant14a}. The fibres feed into the existing AAOmega spectrograph \citep{Sharp06}, a double-beamed spectrograph that offers a range of configurations suitable for galaxy spectroscopy. Since its commissioning in 2012, SAMI observations have been used for studies of galactic winds \citep{Fogarty12,Ho14}, the kinematic morphology--density relation \citep{Fogarty14} and star formation in dwarf galaxies \citep{Richards14}.

The SAMI Galaxy Survey is an ongoing survey to observe $\sim$3400 galaxies covering a wide range of stellar masses, redshifts, and environmental densities. The science goals of the survey are broad, with key themes being the dependence of galaxy evolution on environment, and the build-up of galaxy mass and angular momentum. Observations began in March 2013 and are scheduled to continue until mid-2016.

We present here the Early Data Release (EDR) of the SAMI Galaxy Survey, containing fully calibrated datacubes and associated catalogue information for 107 galaxies selected from the full sample. The EDR sample is described in \mbox{Section\ \ref{sec:sample}}, and a brief overview of the SAMI data reduction process is given in \mbox{Section\ \ref{sec:data_reduction}}. We provide instructions for accessing and using the EDR data in \mbox{Section\ \ref{sec:data_products}}. \mbox{Section\ \ref{sec:quality_metrics}} presents a thorough analysis of the results from the SAMI data reduction pipeline, including key metrics that quantify its current performance. Finally, we summarise our findings in \mbox{Section\ \ref{sec:conclusions}}.

\section{Early data release sample}

\label{sec:sample}

The targets for the SAMI Galaxy Survey are drawn from the GAMA survey G09, G12 and G15 fields, as well as a set of eight galaxy clusters that extend the survey to higher environmental densities. All candidates have known redshifts from GAMA, SDSS or dedicated 2dF observations, allowing us to create a tiered set of volume-limited samples. Full details of the target selection are presented in \citet{Bryant14b}.

The 107 galaxies that form the SAMI Galaxy Survey EDR are those contained in nine fields in the GAMA regions that were observed in March and April 2013. Each field contains 12 galaxies.
One galaxy -- GAMA ID 373248 -- was observed in two separate fields. For this galaxy the EDR contains only the data from the field in which the atmospheric seeing at the time of observation was better (1\farcs2 rather than 2\farcs2).
The full list of galaxies included in the EDR is given in \mbox{Table\ \ref{tab:galaxies}} in the Appendix, which contains detailed information about each galaxy.

\begin{figure}
\includegraphics[width=85mm]{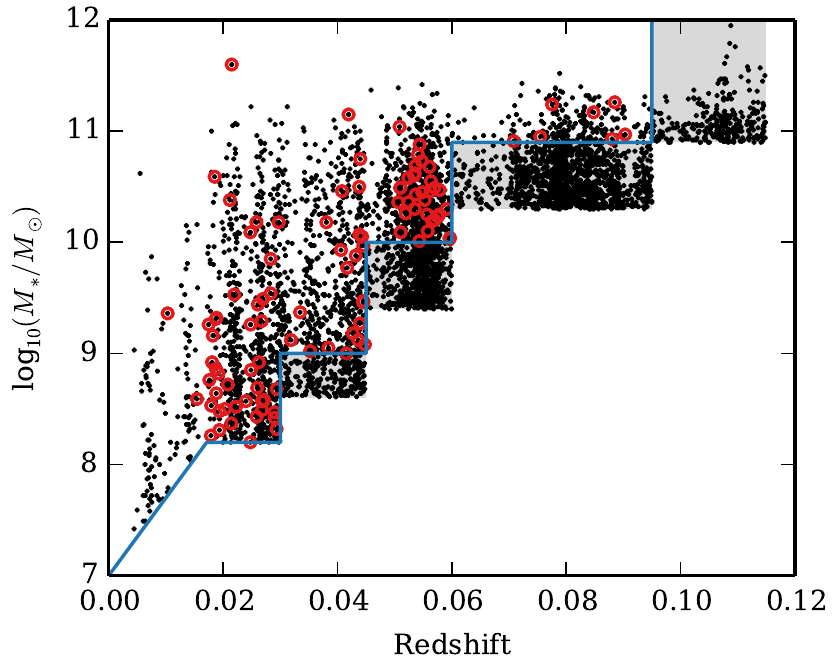}
\caption{Stellar mass and redshift for all galaxies in the field (GAMA) regions of the SAMI Galaxy Survey (black points) and those in the EDR sample (red circles). The blue boundaries indicate the primary selection criteria, while the shaded regions indicate lower-priority targets. Large-scale structure within the GAMA regions is seen in the overdensities of galaxies at particular redshifts.}
\label{fig:mass_vs_z}
\end{figure}

We manually selected the fields in the EDR to provide a representative subsample of the GAMA regions of the full SAMI Galaxy Survey, covering the range of redshifts, stellar masses and galaxy morphologies as completely as possible. The stellar masses \citep{Taylor11,Bryant14b} and redshifts \citep{Driver11} of the EDR sample are shown in the context of the full survey in \mbox{Fig.\ \ref{fig:mass_vs_z}}, which also illustrates the survey's selection criteria. The EDR sample does not contain any galaxies from the high-redshift filler targets, or from the targets with the lowest mass and redshift, resulting in slightly reduced ranges relative to the full sample: $8.2<\log(M_*/{\rm M}_\odot)<11.6$ and $0.01<z<0.09$.

\begin{figure}
\includegraphics[width=85mm]{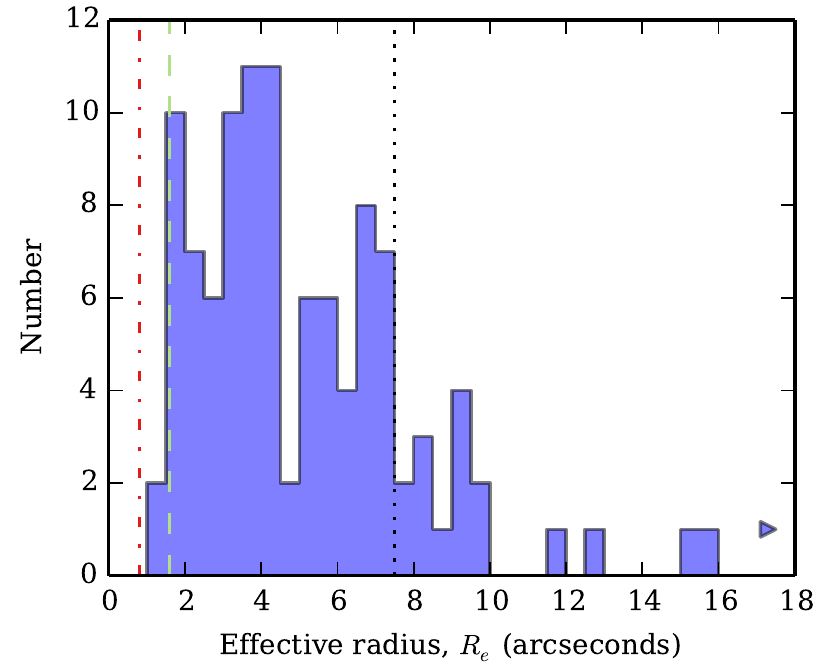}
\caption{Distribution of effective radii in the $r$ band for the galaxies in the SAMI EDR. Vertical lines mark the SAMI fibre radius (dot-dashed red), fibre diameter (dashed green), and hexabundle radius (dotted black). One galaxy, with $R_e=33\farcs05$, is beyond the limits of the plot.
}
\label{fig:re_hist}
\end{figure}

Although no cut was made on the apparent size of the SAMI Galaxy Survey targets, the mass and redshift criteria were chosen with the field of view and fibre size of the SAMI instrument in mind. Most of the galaxies in both the EDR and the full survey are small enough that the field of view reaches at least one effective radius ($R_e$), and large enough that SAMI can resolve their light across a number of fibres. \mbox{Fig.\ \ref{fig:re_hist}} shows the distribution of $r$-band major-axis $R_e$, drawn from the GAMA analysis of SDSS imaging \citep{Kelvin12}, for the galaxies in the EDR. The median $R_e$ is 4\farcs39, with 17 of the 107 having $R_e$ greater than the radius of the SAMI hexabundles (7\farcs5). Only two galaxies have $R_e$ smaller than the fibre diameter (1\farcs6), and none have $R_e$ smaller than the fibre radius.

Details of the nine fields contained in the EDR are given in \mbox{Table\ \ref{tab:edr_fields}}. The table contains the field ID, right ascension and declination of the field centre, date(s) observed, number of exposures, total exposure time, number of galaxies included in the EDR, and full width at half maximum (FWHM) of the output point spread function (PSF; see \mbox{Section\ \ref{sec:fwhm}}). The FWHM is measured at a wavelength of $\lambda_{\rm ref}=5000$\,\AA, and should be scaled by $(\lambda/\lambda_{\rm ref})^{-0.2}$ to apply at other wavelengths.

\begin{table*}
\caption{Fields observed as part of the Early Data Release. See text for a description of the columns. Coordinates are given in decimal degrees at J2000 epoch.}
\label{tab:edr_fields}
\begin{tabular}{lrrlrrrr}
\hline
\multicolumn{1}{c}{Field ID} & \multicolumn{1}{c}{R.A. (deg)} & \multicolumn{1}{c}{Dec.\ (deg)} & \multicolumn{1}{c}{Date(s) observed} & \multicolumn{1}{c}{$N_{\rm exp}$} & \multicolumn{1}{c}{$t_{\rm exp}$\,(s)} & \multicolumn{1}{c}{$N_{\rm gal}$} & \multicolumn{1}{c}{FWHM (\arcsec)} \\
\hline
Y13SAR1\_P003\_09T006 & 140.1079 & $+$1.2923 & 2013 Mar.\ 7 & 7 & 12\,600 & 12 & 1.8 \\ 
Y13SAR1\_P003\_15T008 & 222.6946 & $-$0.3410 & 2013 Mar.\ 7 & 7 & 12\,600 & 12 & 2.1 \\
Y13SAR1\_P005\_09T009 & 131.6677 & $+$2.2979 & 2013 Mar.\ 11 & 7 & 12\,600 & 12 & 2.5 \\
Y13SAR1\_P005\_15T018 & 216.1500 & $-$1.4587 & 2013 Mar.\ 11 & 8 & 14\,400 & 12 & 2.4 \\
Y13SAR1\_P008\_09T013 & 132.5411 & $-$0.0461 & 2013 Apr.\ 14, 16 & 7 & 12\,600 & 12 & 1.8 \\
Y13SAR1\_P009\_09T015 & 139.9145 & $+$0.9335 & 2013 Mar.\ 15 & 7 & 12\,600 & 11 & 2.5 \\
Y13SAR1\_P009\_15T013 & 212.7953 & $-$0.7502 & 2013 Mar.\ 15 & 7 & 12\,600 & 12 & 1.9 \\
Y13SAR1\_P014\_12T001 & 181.0885 & $+$1.7816 & 2013 Apr.\ 12, 13 & 7 & 12\,600 & 12 & 2.2 \\
Y13SAR1\_P014\_15T029 & 214.1435 & $+$0.1883 & 2013 Apr.\ 12 & 7 & 12\,600 & 12 & 2.4 \\
\hline
\end{tabular}
\end{table*}

\section{Data reduction}

\label{sec:data_reduction}

A brief overview of the pipeline used to produce the SAMI Galaxy Survey data products is given below; for a full description, see \citet{Sharp14}. The software itself is publicly available and can be accessed via its listing on the Astrophysics Source Code Library \citep{Allen14}, in which changeset ID \texttt{6c0c801} is the version used for the EDR processing.

\subsection{Extraction of observed spectra}

The first steps of the SAMI pipeline, up to the production of row-stacked spectra (RSS), were carried out using version 5.62 of \textsc{2dfdr}, the standard data reduction software for fibre-fed spectrographs at the Anglo-Australian Telescope (AAT)\footnote{http://www.aao.gov.au/science/software/2dfdr}. \textsc{2dfdr} applies bias and dark subtraction and corrects for pixel-to-pixel sensitivity. Extraction of observed spectra uses an optimal extraction algorithm \citep{Horne86,Sharp10a}, following fibre tramlines identified in separate flat field frames. CuAr arc lamp exposures are used for wavelength calibration. Spectral flat fielding uses a dome lamp system, while the throughput of each fibre is calibrated using the relative strength of sky emission lines (for long exposures) or twilight flat fields (for short exposures). The sky spectrum is measured from 26 dedicated sky fibres, taking a median spectrum after throughput correction, and subtracted from all spectra.

\subsection{Flux calibration and telluric correction}

The RSS data produced by \textsc{2dfdr} were flux calibrated using a combination of spectrophotometric standard stars and secondary standards. As an initial step, all data were corrected for the large-scale (in wavelength) extinction by the atmosphere at Siding Spring Observatory at the observed airmass.

The spectrophotometric standards were in most cases observed on the same night as the galaxies. The transfer function -- ratio of the known stellar spectrum to the observed CCD counts -- was derived accounting for light lost between the fibres. Each transfer function was smoothed before consecutive observations were combined. The RSS data for all galaxy observations were multiplied by the transfer function from the standard observations that were closest in time.

For the telluric absorption correction, secondary standard stars, selected from their colours to be F sub-dwarfs, were observed simultaneously with the galaxies. From their flux-calibrated spectra we derived a correction for absorption in the telluric bands at 6850--6960 and 7130--7360\,\AA. This correction, which accurately describes the atmospheric absorption at the time of observation, was applied to the galaxy spectra in the same exposure to form the fully calibrated RSS frames. The atmospheric dispersion was measured as part of the fit to the secondary standard star, and was then removed when producing the datacube for each galaxy.

\subsection{Formation of datacubes}

\label{sec:cubing}

Each galaxy field was observed in a set of $\sim$7 exposures, with small offsets in the field centres to provide dithering and ensure complete coverage of the field of view. To register each exposure against the others, the galaxy position within each hexabundle was fit with a two-dimensional Gaussian and a simple empirical model describing the telescope offset and atmospheric refraction was fit to the centroids.

Having aligned the individual exposures with each other, we combined all exposures to produce a datacube with regular 0\farcs5 square spaxels. The flux in each output spaxel was taken to be the mean of the flux in each input fibre, weighted by the fractional spatial overlap of that fibre with the spaxel. To regain some of the spatial resolution that would otherwise be lost in convolving the 1\farcs6 fibres with 0\farcs5 spaxels, the overlaps were calculated using a fibre footprint with only 0\farcs8 diameter (a drizzle-like process, see \citealt{Sharp14,Fruchter02}). The variance information was fully propagated through the cubing process.

Because each input fibre overlaps with more than one output spaxel, the flux measurements in the datacubes are covariant with nearby spaxels.  This is a generic issue for any data that is resampled that is resampled onto a grid; for example, \citet{husemann13} discuss the problem in the context of the CALIFA survey.  A crucial consequence is that, when spectra from two or more spaxels are summed, the variance of the summed spectrum is not equal to the sum of the variances of the individual spectra. Similarly, a model fit to the datacube would need to account for the covariance between spaxels. The format in which the covariance information is stored is described in \mbox{Section\ \ref{sec:covariance}} (further details are given in \citealt{Sharp14}) and its effect is quantified in \mbox{Section\ \ref{sec:covariance_effect}}.

The earlier flux calibration step assumed that the atmospheric conditions did not vary between the observations of the galaxies and the spectrophotometric standard star. Although this is usually a good approximation, the atmospheric transmission can vary during the night. 
These variations were measured by fitting a Moffat profile to the datacube of the secondary standard star, extracting its full spectrum, then integrating across the SDSS $g$ band to find the observed flux.
All objects in the field were then scaled by the ratio of the true flux of the star -- from the SDSS or, for some cluster fields, VLT Survey Telescope (VST) ATLAS \citep{Shanks13} imaging catalogues -- to the observed flux, under the assumption that the atmospheric variation is the same across the entire field.

\section{Data products}

\label{sec:data_products}

\begin{figure*}
\includegraphics[width=175mm]{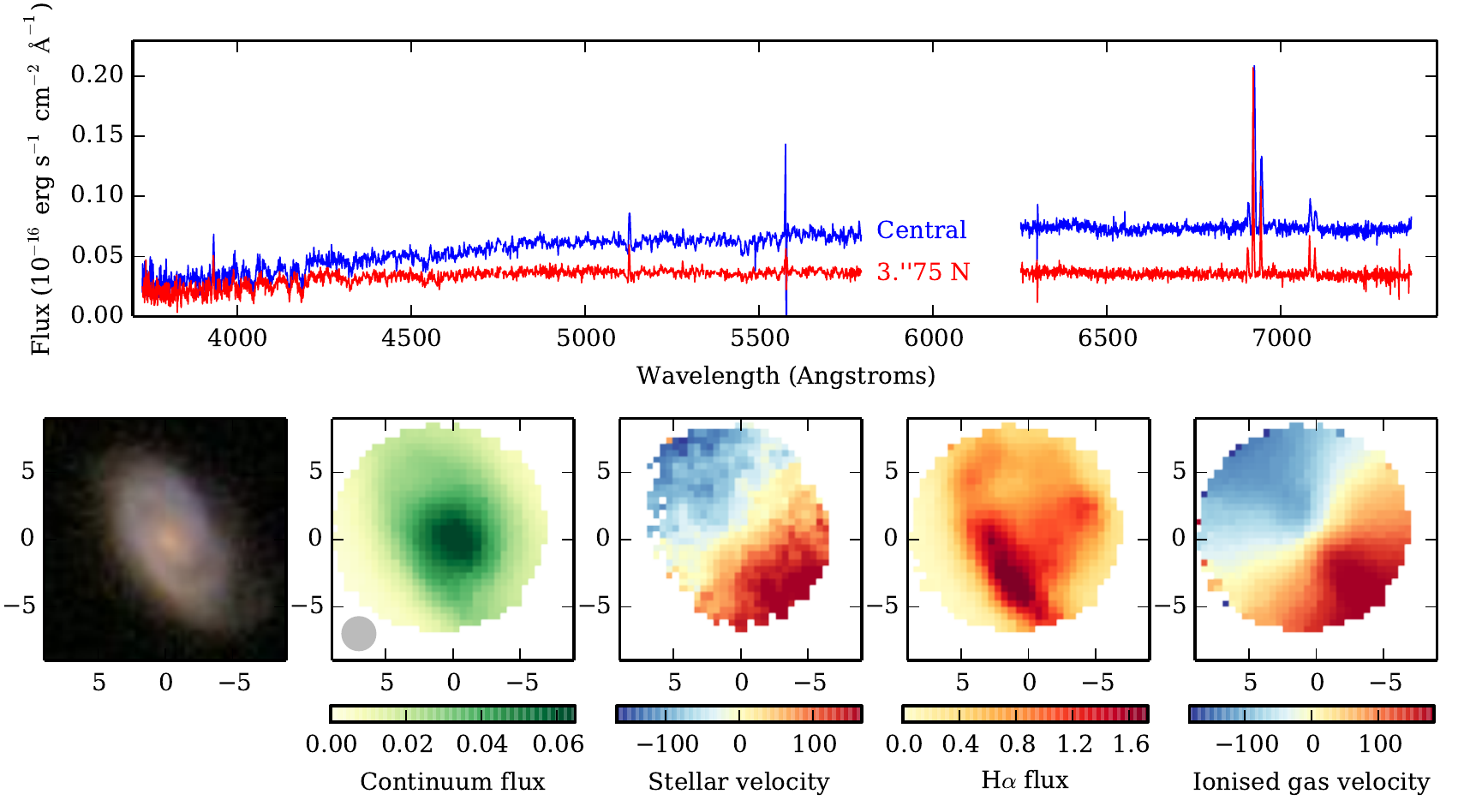}
\caption{Example SAMI data for the galaxy 511867, with $z=0.05523$ and $M_*=10^{10.68}{\rm M}_\odot$. Upper panel: flux for a central spaxel (blue) and one 3\farcs75 to the North (red). Lower panels, from left to right: SDSS $gri$ image; continuum flux map ($10^{-16}$\,erg\,s$^{-1}$\,cm$^{-2}$\,\AA$^{-1}$); stellar velocity field (km\,s$^{-1}$); H$\alpha$ flux map ($10^{-16}$\,erg\,s$^{-1}$\,cm$^{-2}$); H$\alpha$ velocity field (km\,s$^{-1}$). The two velocity fields are each scaled individually. For the stellar velocity map, only spaxels with per-pixel signal-to-noise ratio $>$5 in the continuum are included. Each panel is 18\arcsec\ square, with North up and East to the left. The grey circle in the second panel shows the FWHM of the PSF.
}
\label{fig:example_1}
\end{figure*}

\begin{figure*}
\includegraphics[width=175mm]{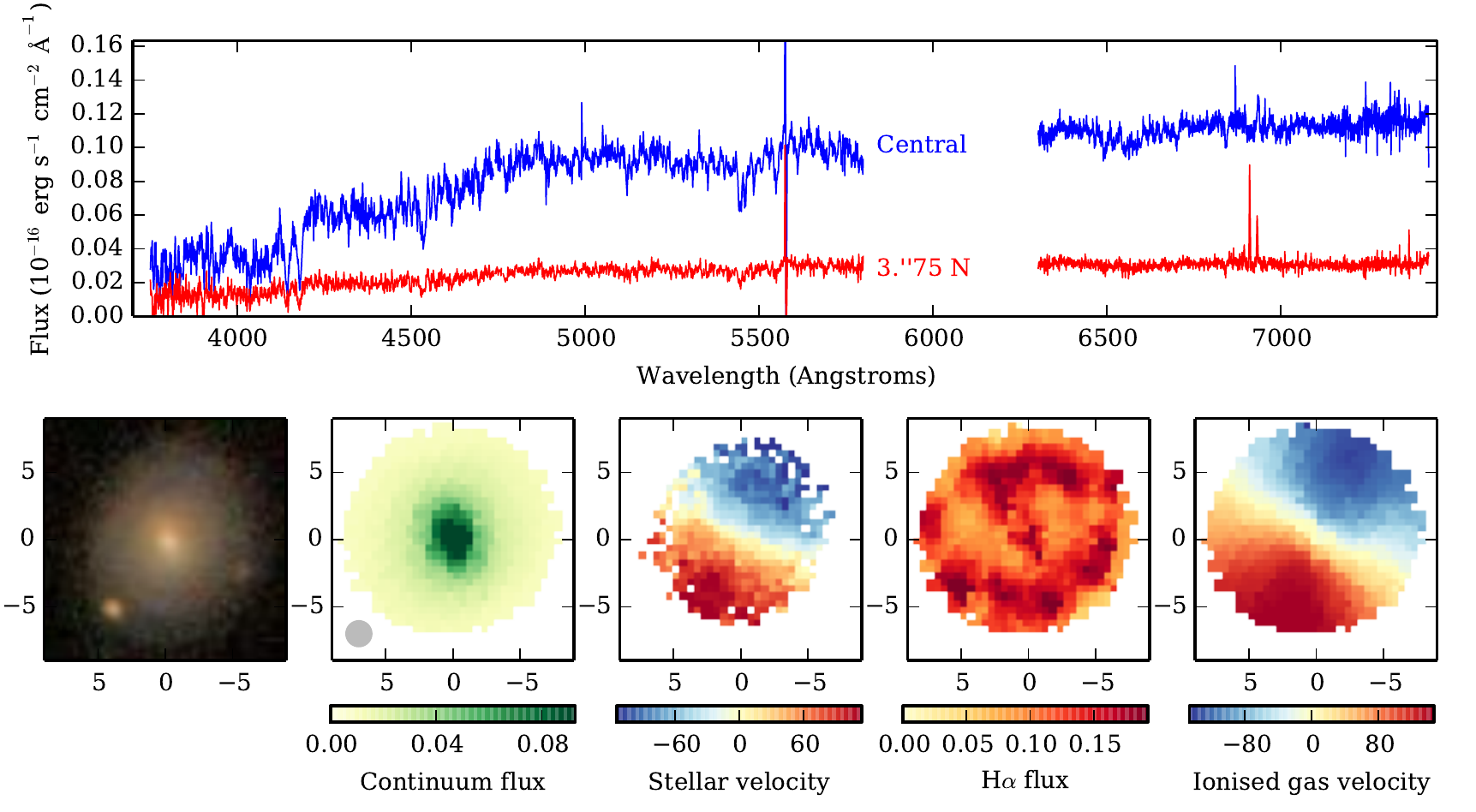}
\caption{As \mbox{Fig.\ \ref{fig:example_1}}, for the galaxy 599761 with $z=0.05333$ and $M_*=10^{10.88}{\rm M}_\odot$.}
\label{fig:example_2}
\end{figure*}

Examples of SAMI data are shown in \mbox{Figs.\ \ref{fig:example_1}} and \ref{fig:example_2}, illustrating the nature and quality of the available data. For each of two galaxies from the EDR sample we show example spectra from individual spaxels, along with maps of the continuum flux, stellar velocity, H$\alpha$ flux and H$\alpha$ velocity. 

A visual inspection indicates that SAMI produces high-quality spectra across the visible extent of each galaxy, with parameters such as stellar velocity fields and emission-line maps readily measurable in almost all cases. We quantify the performance of the data reduction pipeline and the quality of the data products in \mbox{Section\ \ref{sec:quality_metrics}}. In the following subsections we describe the details of the data products and how to access them.

\subsection{File format}

The final product of the data reduction pipeline for each galaxy is a pair of datacubes, covering the wavelength ranges of each arm of the AAOmega spectrograph. For the SAMI Galaxy Survey observations, AAOmega is configured to give a spectral resolution of $\sim$1700 ($\sim$4500) and wavelength range of 3700--5700\,\AA\ (6300--7400\,\AA) in the blue (red) arm. Each of the datacubes is provided as a FITS file \citep{Pence10} with multiple extensions, namely:
\begin{itemize}
\item \texttt{PRIMARY}: Flux datacube
\item \texttt{VARIANCE}: Variance datacube
\item \texttt{WEIGHT}: Weight datacube
\item \texttt{COVAR}: Covariance between related pixels
\end{itemize}
The flux is in units of 10$^{-16}$\,erg\,s$^{-1}$\,cm$^{-2}$\,\AA$^{-1}$\,spaxel$^{-1}$, and the variance in the same units squared. The weight datacube describes the relative exposure of each spaxel [see \citet{Sharp14} for details] and is used in applying the covariance information (see \mbox{Section \ref{sec:covariance}}).

Each of the flux, variance and weight datacubes has dimensions 50$\times$50$\times$2048, i.e.\ 50$\times$50 spaxels (each spaxel 0\farcs5 square) and 2048 wavelength slices (1.04/0.57\,\AA\ sampling in the blue/red datacubes). Spaxels with known problems, such as those affected by cosmic rays or bad columns on the CCDs, are recorded as `not a number' (NaN) in all extensions. Measurements of the spatial FWHM in each field are given in \mbox{Table \ref{tab:edr_fields}}, as derived from the fits to the standard stars made during flux calibration.

\subsection{Covariance}

\label{sec:covariance}

The variance array is correctly scaled for analysis of the spectrum in each individual spaxel. However, due to the re-sampling procedure by which the datacubes were derived, the increase in signal-to-noise ratio (S/N) from combining adjacent spaxels is lower than a naive propagation of the variance would suggest. The information required to reconstruct the correct variance of a summed spectrum is encoded in the covariance array, which is recorded in a compressed form in the \texttt{COVAR} extension.

The covariance is stored as a 50$\times$50$\times$5$\times$5$\times$$N$ array, where $N$ is the number of wavelength slices for which the covariance has been calculated and stored. For each wavelength slice $m$, the value of \texttt{COVAR}[$i,j,k,l,m$] records the \textit{relative} covariance between the spaxel $(i,j)$ and one nearby spaxel, for data that have been weighted by the relative exposure times. This value should be multiplied by the weighted variance of the $(i,j)^{\rm th}$ spaxel to recover the true covariance. The central value in the $k$ and $l$ directions ($k=2$, $l=2$ for 0-indexed arrays) corresponds to the relative covariance of the $(i,j)^{\rm th}$ spaxel with itself, which is always equal to unity. Surrounding values give the relative covariance of the $(i,j)^{\rm th}$ spaxel with all locations up to two spaxels away. For example,
\begin{equation*}
\texttt{COVAR}[i,j,1,-1,m]\times\texttt{VARIANCE}[i,j,m^\prime]\times\texttt{WEIGHT}[i,j,m^\prime]^2
\end{equation*}
gives the covariance between the weighted data points
\begin{equation*}
\texttt{PRIMARY}[i,j,m^\prime]\times\texttt{WEIGHT}[i,j,m^\prime]
\end{equation*}
and
\begin{equation*}
\texttt{PRIMARY}[i+1,j-1,m^\prime]\times\texttt{WEIGHT}[i+1,j-1,m^\prime],
\end{equation*}
where $m^\prime$ is the wavelength pixel corresponding to the covariance slice $m$ (see below). At separations larger than two spaxels the covariance is negligible.

The positions of the wavelength slices at which the covariance was calculated were chosen to accurately sample the regions where the covariance changes most rapidly, which occurs where the drizzle locations were recalculated due to atmospheric dispersion. The number of such slices is given by the \texttt{COVAR\_N} header value in the \texttt{COVAR} extension, and the location of each is given by the header values \texttt{COVARLOC\_1}, \texttt{COVARLOC\_2} etc., recorded as pixel values. Relative covariance values at intervening positions can be inferred by copying the values at the nearest wavelength slice provided (the difference between copying and interpolating is negligible relative to the covariance values).

\subsection{Data access}

\label{sec:data_access}

\begin{figure*}
   \centering
   \includegraphics[width=16cm]{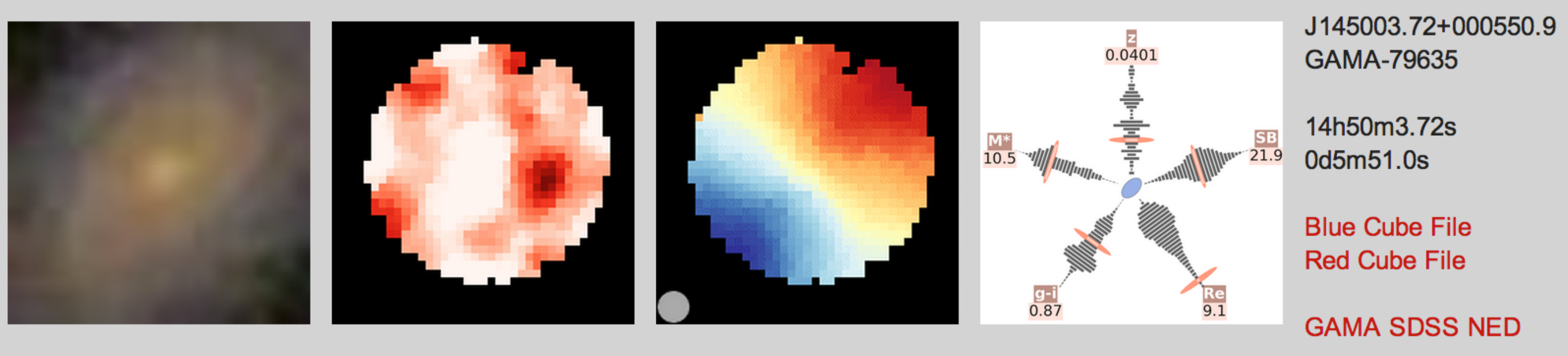}
   \caption{Example representation on the SAMI Galaxy Survey website of a galaxy (GAMA ID 79635) with strong H$\alpha$ emission. From left to right, the panels are: SDSS $gri$ image of the galaxy; logarithmically-scaled H$\alpha$ flux map; H$\alpha$ velocity map; `starfish diagram' placing the galaxy's properties within the context of the full sample; the galaxy's coordinates and direct links to SAMI datacubes and ancillary data. The grey circle in the third panel shows the FWHM of the PSF (see \mbox{Section\ \ref{sec:fwhm}}). The arms of the starfish show the galaxy's redshift, surface brightness at the effective radius (mag\,arcsec$^{-2}$), effective radius (arcsec), $g-i$ colour, and stellar mass (log solar masses). The galaxy's value is marked by a red line, while the underlying grey histogram shows the distribution of the property for the full sample. For all arms the physical value increases as you move away from the centre of the starfish. The H$\alpha$ flux and velocity are derived from a single Gaussian profile fit to the H$\alpha$ emission line. The small ellipse in the centre shows the ellipticity and position angle of the galaxy.}
   \label{fig:thumbnail_example_1}
\end{figure*}

\begin{figure*}
   \centering
   \includegraphics[width=16cm]{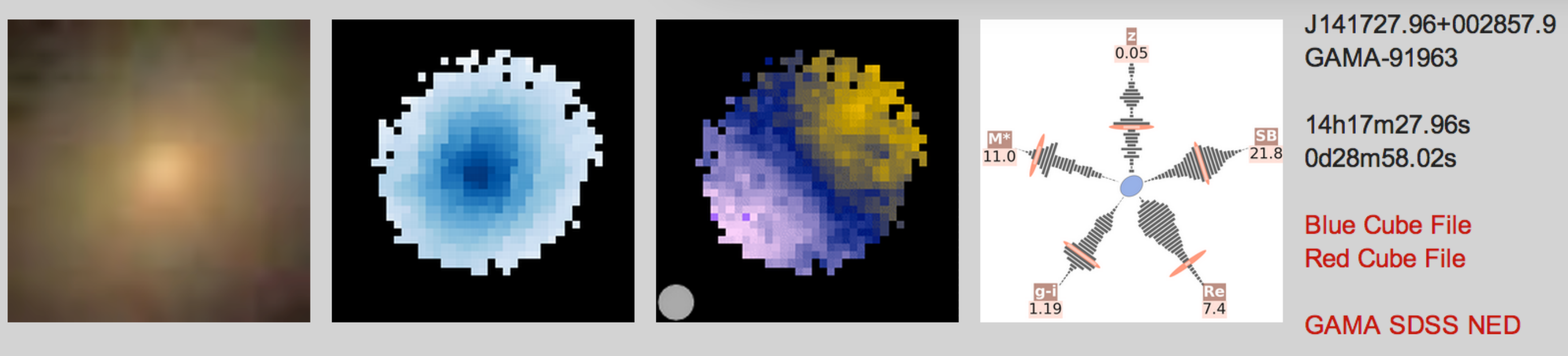}
   \caption{As \mbox{Fig.\ \ref{fig:thumbnail_example_1}}, for a galaxy with strong stellar continuum (GAMA ID 91963). In this case the second and third panels show the integrated light with logarithmic scaling and the stellar velocity field, respectively. The velocity field is extracted using \textsc{ppxf} \citep{Cappellari04}, and is shown for all spaxels with S/N$>$5 in the continuum.}
   \label{fig:thumbnail_example_2}
\end{figure*}

The datacubes for the 107 galaxies in the EDR sample can be downloaded from the SAMI Galaxy Survey website\footnote{http://sami-survey.org/}. A table of data from previous imaging and spectroscopic observations is also provided; we describe the table contents in the appendix. For each galaxy the website also shows the SDSS $gri$ thumbnail images, maps of the flux and velocity for either H$\alpha$ or the stellar continuum, and a `starfish diagram' giving a visual representation of the key parameters of the galaxy \citep{Konstantopoulos14}; two examples are shown in \mbox{Figs.\ \ref{fig:thumbnail_example_1}} and \ref{fig:thumbnail_example_2}.

\section{Quality metrics}

\label{sec:quality_metrics}

In this section we present several tests to assess data reduction quality at all stages in the SAMI pipeline. For full details of the processing steps we refer the reader to the description of the SAMI data reduction pipeline in \citet{Sharp14}. Our aim here is to allow the reader to assess the quality of the current SAMI EDR reductions and also to point out any limitations in the current processing. In order to maximise the statistical power of the measurements, results are calculated for all SAMI Galaxy Survey observations up to May 2014, not just the EDR fields, except where noted.

\subsection{Preliminary reductions}

First we assess the quality of the steps up to the production of wavelength-calibrated, sky-subtracted, row-stacked spectra.  All of the processing up to this stage is carried out using the {\sc 2dfdr} data reduction pipeline, developed by the Australian Astronomical Observatory (AAO). Each reduction step impacts on the accuracy of sky-subtraction \citep{Sharp10c} and our ability to accurately flux calibrate SAMI data.  Below we assess the quality of key steps in the processing in the order they are carried out.

\subsubsection{Cross-talk between fibres at the CCD}

Here we consider only cross-talk between overlapping PSFs on the detector, rather than between the fused fibres within the hexabundle itself (which is generally minimal compared to that generated by atmospheric seeing, see \citealt{Bryant11} for a discussion of this effect). SAMI has 819 fibres (including 26 sky fibres) which are distributed across the 4096 spatial pixels of the AAOmega spectrograph. At the slit, the 61 fibres from each IFU are arranged in a contiguous section, bounded by two sky fibres. Thirteen of such sections are placed end-to-end to form the complete slit. Within an IFU, adjacent fibres on the slit are typically next-but-one neighbours in the IFU, although the circular close packing of the SAMI IFUs results in some exceptions to this rule.

The typical separation between fibres on the CCD is $\simeq4.75$ pixels, and the width of the fibre profiles is $\simeq2.8$ pixels (FWHM), so there is significant overlap between fibres.  A simple window of width 4.75 pixels centred on a fibre will typically contain $\simeq95$ per cent of the flux from that fibre, with the rest falling into adjacent windows. With a perfect model of the PSF, the optimal extraction technique, which simultaneously fits multiple overlapping profiles, would correct for any overlapping flux.

We examine the ability of the current pipeline to correct for cross-talk using observations of bright spectrophotometric flux standards, and comparing the flux in fibres which are adjacent on the slit, but not at the IFU.  Comparing the residual flux in fibres adjacent to bright stars on the slit we find that typically $\sim0.5$ per cent of the flux is incorrectly assigned to the adjacent fibres.  The relative effect of this additional flux depends linearly on the flux ratio of the two fibres. This is largely due to the wings of the fibre PSF being broader than the current simple Gaussian model used in extraction. Future improvements to the pipeline will introduce more accurate model PSFs to reduce the cross-talk.

\subsubsection{Flat field accuracy}

Fibre flat field frames are used both to trace the fibre locations across the detector and to calibrate the fibre-to-fibre relative chromatic response.  Errors in flat fielding eventually propagate to sky subtraction and flux calibration systematics, so careful attention has to be paid to flat field accuracy.

In order to characterize our flat field accuracy, we apply the flat fields to twilight sky observations and then measure the uniformity in colour of these frames.  In the test below we will compare three different cases:  i) The default dome flat fields used to calibrate SAMI observations; ii) flat fields using a lamp illuminating the field of view from a position on a flap approximately $\sim1$\,m in front of the prime focus corrector; iii) no flat fields applied.  We correct for overall fibre transmission and median smooth the twilight sky spectrum (to reduce the impact of shot noise on our measurement).  We then divide each fibre by the median twilight sky spectrum.

\begin{figure*}
\includegraphics[width=85mm]{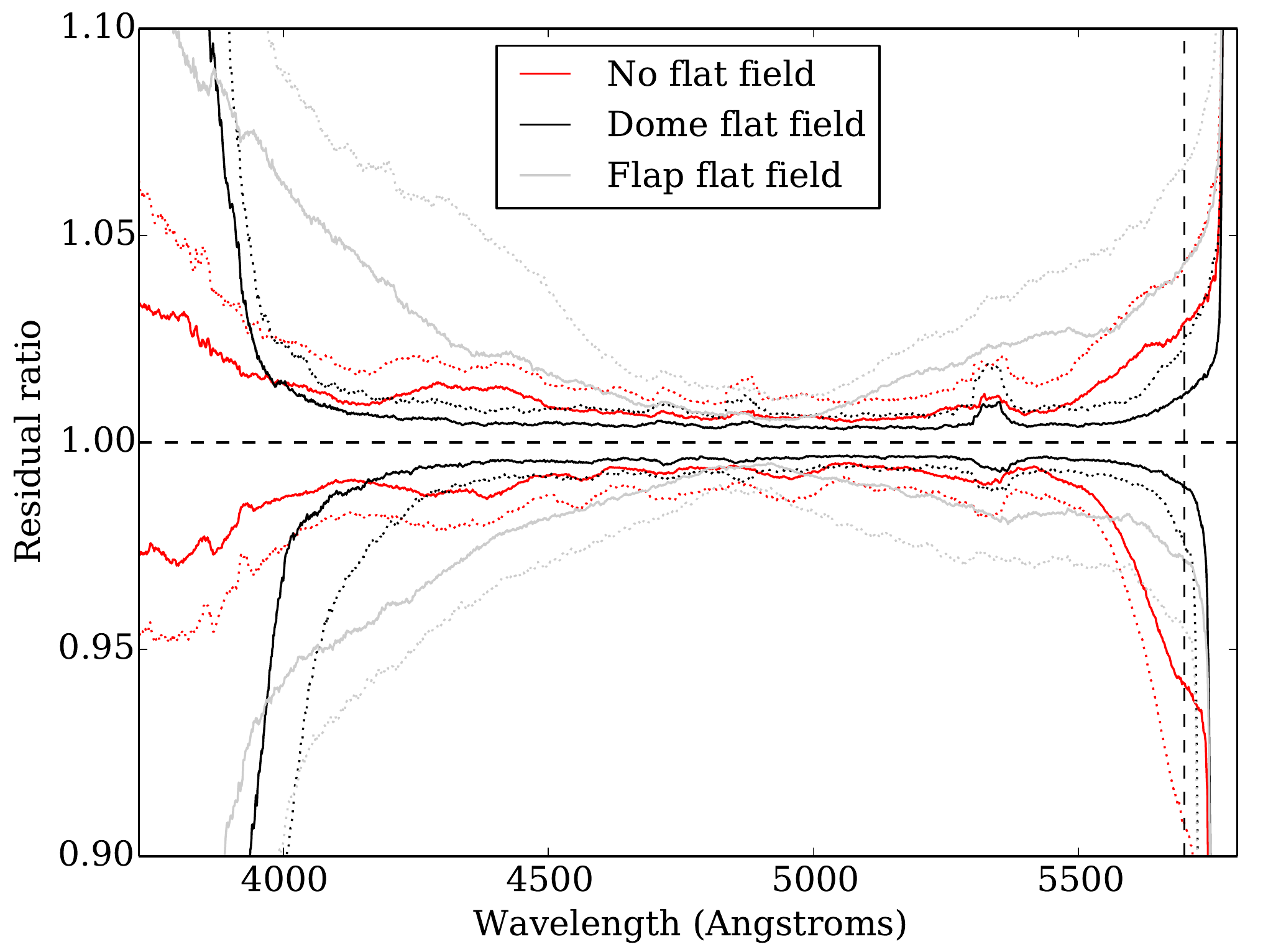}\hspace{5mm}\includegraphics[width=85mm]{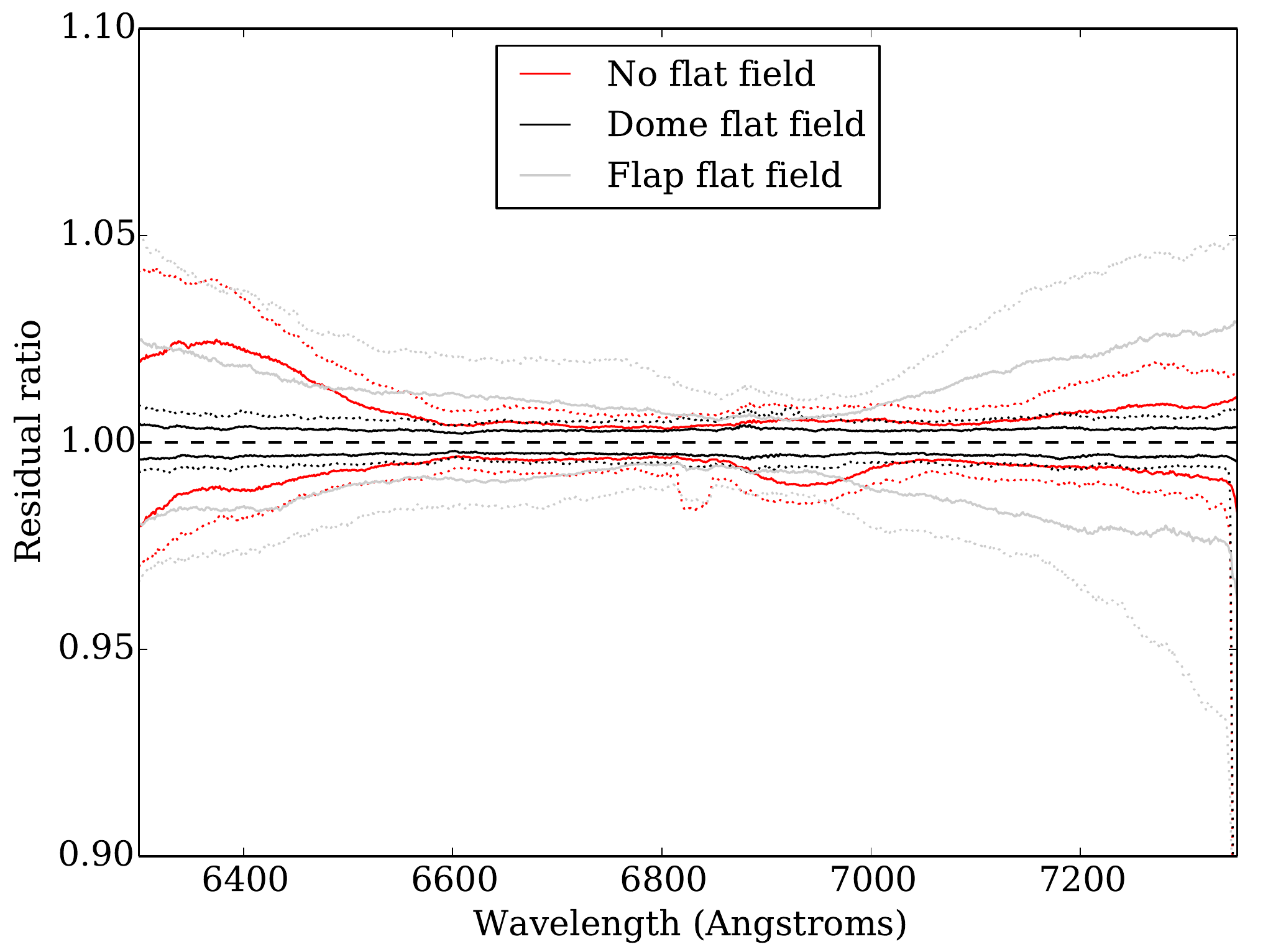}
\caption{Comparison of the scatter in calibrated twilight sky colours for the SAMI blue (left) and red (right) arms.  In each plot we show the 68th percentile range (i.e.\ $1\sigma$, solid lines) and 95th percentile range (i.e.\ $2\sigma$, dotted lines) for the scatter between the 819 SAMI fibres as a function of wavelength.  Red lines show the variation present if no flat fielding is applied.  The black lines are applying dome flats and the grey lines are applying the flap flats.  The vertical dashed line shows the location of the dichroic cut off at 5700\,\AA.}
\label{fig:ff_errors}
\end{figure*}

In \mbox{Fig.\ \ref{fig:ff_errors}} we plot the 68th (solid lines) and 95th (dotted lines) percentiles of the fibre-to-fibre scatter as a function of wavelength.  The red lines show the native variation in fibre colour response without any flat fielding applied.  This is typically $\sim1$--2 per cent (for 68th percentile range) or less in both red and blue arms.  Applying the dome flat field (black lines) improves the colour uniformity to 0.5--1 per cent in the blue arm and 0.3--0.4 per cent in the red arm.

The exception to this is at the far blue end of the blue arm, where applying the dome flat field makes the agreement {\it worse}, with 68th-percentile ranges rapidly rising above 10 per cent. The reason for this is stray light on the detector at the blue end, which is a significant fraction of the signal in the far blue for the flats. The stray light components are not dispersed spectrally so contain significant red light, although they land on the blue end of the detector. They are only significant for the flat fields, which have much lower counts in the blue than the red (as is often the case for spectral flat fields, but not for most astronomical targets).

Efforts to empirically model the stray light in the AAOmega spectrograph are ongoing, but our current proceedure is to fibre flat field the blue arm data from spectral pixels 500 to 2048, but not apply any fibre flat fielding at pixels less than 500 (4245\,\AA).  This results in a maximum error in fibre colour response of $\sim 2$ per cent at wavelengths below 4000\,\AA.  In contrast to the blue arm, flat fielding in the red arm is straightforward.

Although not used in SAMI reductions, for comparison we show the results of using the `flap' flat field lamp (grey lines in \mbox{Fig.\ \ref{fig:ff_errors}}).  This results in poorer colour flat fielding than using no flat fielding at all.  The cause of this is the diverging beam of this flat lamp at the SAMI focal plane which over-fills the fibres and causes the output beam angle of the fibre slit to overfill the spectrograph collimator.  This highlights the need to feed calibration sources to the instrument at an appropriate f-ratio.

\subsubsection{Wavelength Calibration}

Wavelength calibration is carried out in the usual manner, using CuAr arc lamps with known spectral features to derive a transformation from pixel coordinates to wavelength coordinates (air wavelengths).  Arc lamps are located at a position in front of the prime focus corrector, so as for the `flap' flat field above, there is a difference between the input beam coming from the sky and that from the arc lamps.  This results in small but significant changes in the measured spectral PSF and can cause an offset between the true and measured wavelength scale of $\simeq0.1$ pixels.

%
%
\begin{figure*}
\includegraphics[width=175mm]{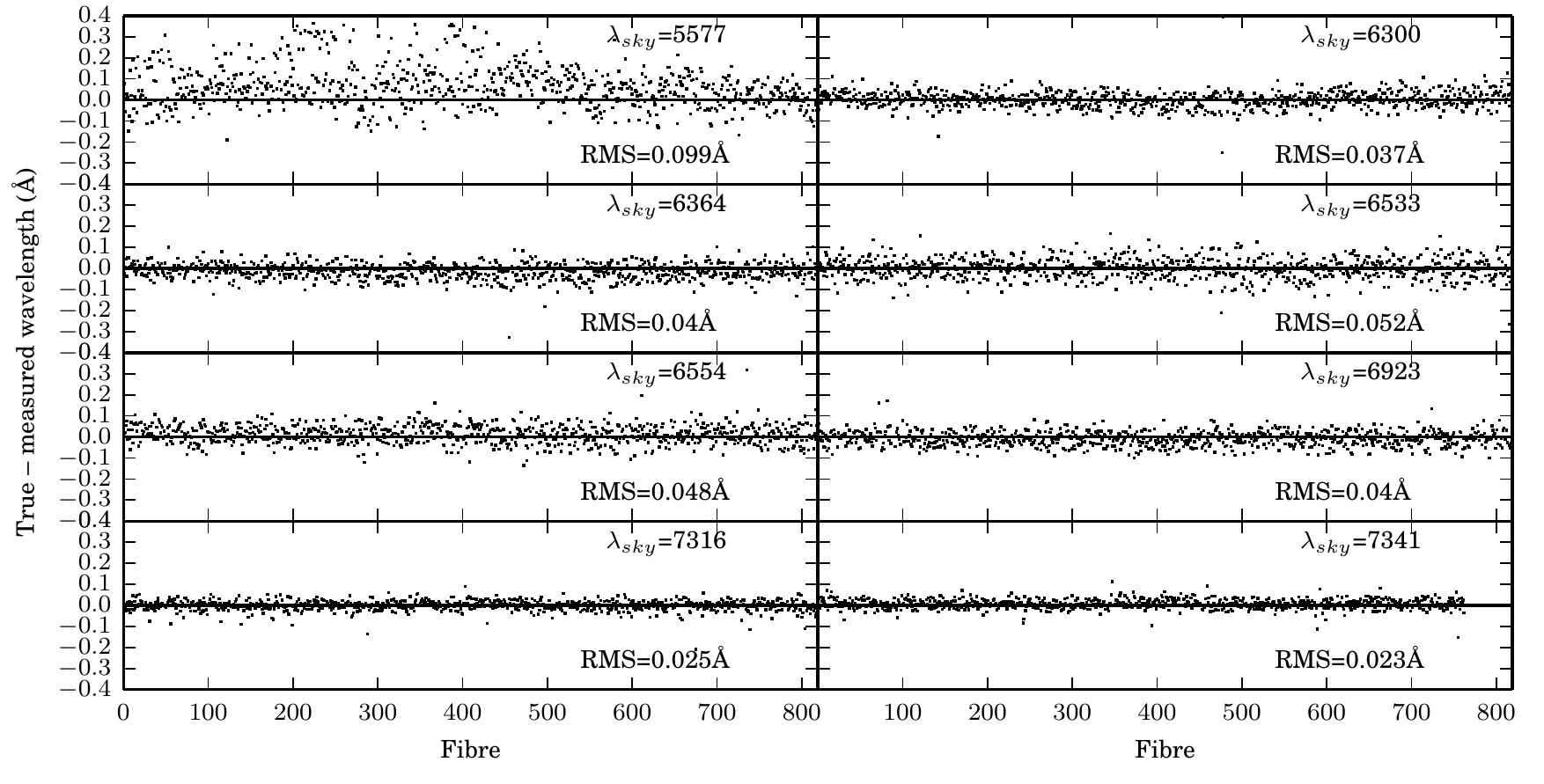}
\caption{The difference between true and measured night-sky emission-line wavelength as a function of fibre number for different strong sky lines in the SAMI wavelength range, as measured from a typical 1800s SAMI galaxy exposure.  The top left plot is for the 5577-\AA\ line, which is the only one in the blue arm.  This line shows larger scatter due to the lower instrumental resolution, and a small mean residual offset.  The other lines are all in the red arm and show smaller scatter (per \AA) due to the higher resolution.  The cases where a number of fibres do not show a measurement (e.g. 7341\,\AA) are caused by a combination of bad columns on the CCD or the edge of the detector combined with spectral curvature across the CCD.  Occasional outliers are due to cosmic rays influencing the fits on individual line positions.}
\label{fig:skywavelength}
\end{figure*}

To quantify and correct this we apply a secondary wavelength calibration using the night sky emission lines. The correction is not a simple linear shift, but has a quadratic variation which depends on the position across the CCD. This can be applied successfully in the red arm, where there are a number of sky lines.  In the blue arm, which only contains the 5577-\AA\ sky emission line, we cannot succesfully apply the correction.

The relative position after correction of the night sky lines with respect to their expected wavelength is shown in \mbox{Fig.\ \ref{fig:skywavelength}} for a typical 1800s galaxy exposure as a function of fibre number.  It can be seen that systematic differences for the sky lines in the red arm are negligible (after applying secondary calibration using the sky lines), with typical root mean square (RMS) scatter of $\simeq0.02$--0.06\,\AA, corresponding to $\simeq0.035$--0.11 pixels.  In the blue arm the RMS is larger (as expected given the lower resolution) and there is a weak $\simeq0.05$--0.1 pixel systematic offset between the true and expected position of the 5577-\AA\ line due to the distortion discussed above.  The RMS (from zero offset) is $\simeq0.1$\,\AA\ (or $\simeq0.1$ pixels).

In summary, the wavelength calibration across all SAMI data is currently accurate to $\simeq0.1$ pixels or better (RMS).  Further developments to use the full optical model to fit a more accurate and robust wavelength solution [as described in \cite{Childress14} for the WiFeS instrument], as well as matching directly to the sky using twilight sky observations will be implemented in future data releases.

\subsubsection{Fibre throughput calibration}

%

As described above (and discussed in detail in \citealt{Sharp14}), the relative normalization of fibre throughputs in SAMI is carried out based on the integrated flux in the night sky lines (twilight sky observations are not consistently available, and the spatial flatness of the dome flats has not been quantified).  This has a direct influence on our ability to carry out precise sky subtraction \citep{Sharp13}.  To examine the thoughput estimates we measure the RMS scatter in throughput over the typical $7\times1800$s exposures taken for each field.  Within each set of observations the throughputs have a typical RMS scatter (median RMS over all fibres) of 0.7--0.9 per cent for both red and blue arms.  This is consistent with the scatter being largely due to the photon counting noise of the flux in the sky lines.

If we look at the throughput across observations of different fields, the RMS increases to between 2 and 4 per cent.  This is due to small amounts of differential focal ratio degradation between the different plate configurations (with IFU bundles being placed at different locations within the field of view).  The increased scatter is unlikely to be related to variability in the sky lines, as this would require gradients across the field of view that were stable during the 4~hours for which each field was observed, despite the motion of the telescope, but then changed between fields. As we calibrate the fibre throughputs for each field independently, we can achieve throughput calibration uncertainties that are less than 1 per cent. 

\subsubsection{Sky subtraction accuracy}

%
%

%
%
%
%
%

Next we examine the accuracy with which we can subtract the night sky emission from our data.  Of particular importance is the subtraction of sky continuum emission.  Errors in continuum subtraction can be particularly serious when binning contiguous regions of IFU data to achieve good signal-to-noise (S/N) in regions of low surface brightness (i.e. the outer parts of galaxies).  Therefore we largely focus on continuum sky residuals, noting that because the throughput calibration is based on the emission lines, they will by design have small residual flux after sky subtraction (although limitations in resampling and wavelength calibration can continue to cause emission line residuals larger than expected from Poisson statistics).

\begin{figure}
\includegraphics[width=85mm]{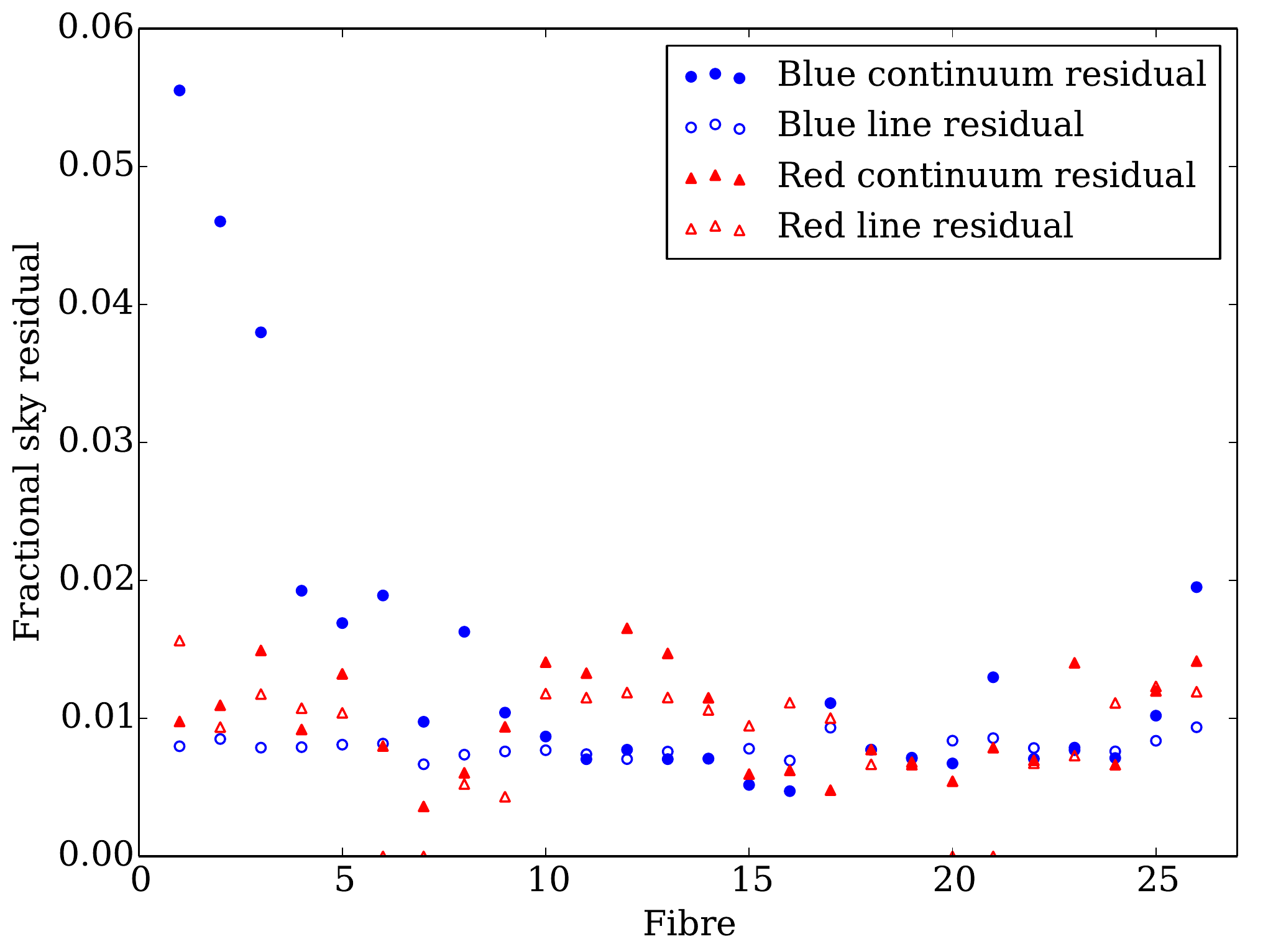}
\caption{The median sky subtraction residual for continuum (filled symbols) and emission lines (open symbols) for the blue (circles) and red (triangles) arms of the SAMI data.  The residuals are plotted for each of the 26 sky fibres in the SAMI instrument.}
\label{fig:skysub}
\end{figure}

To assess the accuracy of the sky subtraction we look at data from the 26 sky fibres in the SAMI instrument and quantify the fractional flux remaining in the spectra from these fibres after sky subtraction.  For each fibre we then calculate the median fractional sky subtraction residual over 367 frames (all the data taken in 2013).  The results for blue and red CCDs are shown in \mbox{Fig.\ \ref{fig:skysub}}.  We first note that the emission line residuals (open symbols) are uniformly low for both the blue and red arms.  The median and 90th percentile fractional residual line fluxes are 0.8, 2.0 per cent (blue arm) and 0.9, 2.9 per cent (red arm).  The blue arm has lower residuals simply due to the fact that only a single emission line (the 5577-\AA\ line) is used for both throughput and for assessing the sky subtraction residuals, allowing apparently sub-Poisson precision in the removal of the line.  

The continuum residuals (filled symbols in \mbox{Fig.\ \ref{fig:skysub}}) show coherent structure along the SAMI slit (from sky fibre 1 to 26).  In particular, the blue arm has regions with up to 5 per cent sky residuals, but overall the median continuum sky subtraction is at the level of 1.2 per cent, with a 90th percentile range of 4.6 per cent.  Similar features in the continuum sky subtraction residuals are seen in the red arm with a median of 0.9 per cent and 90th percentile of 3.1 per cent.  We have examined a set of SAMI observations targeting blank fields to examine whether there is any systematic difference between sky fibres and IFU fibres, and find that the sky subtraction residuals are the same for both. Changes in focal ratio degradation may be affecting the sky subtraction accuracy, but after the throughput correction only chromatic variations remain, which are typically at a smaller level than the observed residuals.

It is worth noting that the continuum residuals (filled symbols) are not correlated with the emission line residuals. This along with the coherent variation along the slit suggests that the continuum sky subtraction residual seen here is due to incomplete modelling of the large-scale stray light across the detector as well as any deviations of the spatial PSF (the cross-section of the fibre's light along the slit direction) from its assumed Gaussian profile, both of which cause systematic errors in the extracted flux. We aim to correct both of these in the near future with more accurate empirical models for stray light and the fibre spatial PSF.

\subsection{Flux calibration}

As described in Section \ref{sec:data_reduction}, the primary flux calibration of SAMI data relies on separate observations of spectrophotometric standard stars, typically observed on the same night as the target galaxies. The nature of the flux calibration procedure allows two forms of quality control metrics to be derived: internal self-consistency checks between calibrations taken at different times, and comparisons to external photometric data. Both are described in the following subsections (\ref{sec:fluxcal_stability} -- \ref{sec:fluxcal_zero}), before measurements of the accuracy of the telluric correction (\ref{sec:telluric}).

\subsubsection{Stability of flux calibration}

\label{sec:fluxcal_stability}

With multiple observations of standard stars over a period of more than a year, we are able to quantify the stability of the flux calibration over both short and long timescales. Comparing the measured transfer function between different observations, we find a moderately large scatter of 11.1 per cent in the normalisation, due primarily to variations in atmospheric transmission. To quantify the time-variation of the transfer function as a function of wavelength we first remove the changes in normalisation, by dividing each transfer function by its mean value. Then, we take the standard deviation of normalised values at each wavelength pixel, $\sigma_f(\lambda)$, and divide by the mean normalised value at that wavelength, $\bar{f}(\lambda)$.

\begin{figure}
\includegraphics[width=85mm]{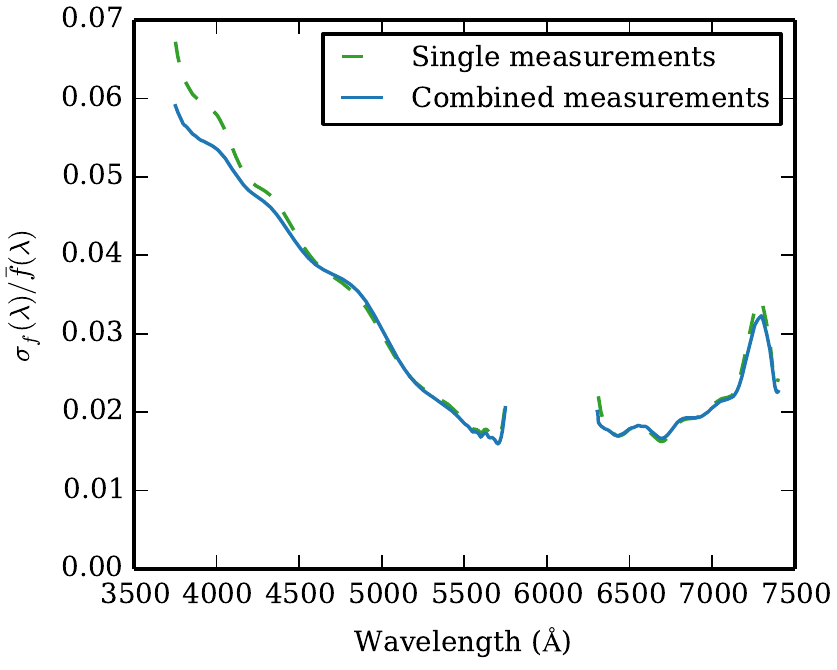}
\caption{One-sigma fractional scatter in the measured flux calibration transfer functions, over all observations to date. The scatter is plotted as a fraction of the mean value across the same observations. }
\label{fig:fluxcal_stability}
\end{figure}

The resulting fractional scatter is plotted in \mbox{Fig.\ \ref{fig:fluxcal_stability}}. The scatter is shown both for the combined measurements used to calibrate the SAMI data, and the individual observations from which they were derived. Combining the measurements reduces the scatter where it is intrinsically the highest, for wavelengths $<4500$\,\AA\ and in the telluric feature at $\sim7200$\,\AA. At all wavelengths the final scatter is below 6 per cent, and below 4 per cent for wavelengths longer than 4500\,\AA.

The observed scatter is likely a combination of time variation in the end-to-end throughput as a function of wavelength, and limitations in the extraction procedure for point sources. When considering individual observing runs -- between 4 and 13 nights of contiguous data -- the observed scatter is typically reduced by $\sim30$ per cent, indicating that at least some of the variation is due to small changes in the system throughput (e.g.\ primary mirror re-aluminzation, slit alignment) or atmospheric conditions (e.g.\ dust/aerosol content).

\subsubsection{Accuracy of colour correction}

The accuracy of the flux calibration procedure can be seen by comparing the inferred colours of the secondary standard stars to the colours measured from previous imaging surveys, either SDSS or VST-ATLAS. There are two stages at which this can be done: for the individual flux-calibrated RSS frames, and for the final datacubes.

In each case, an extracted spectrum representing the complete flux of the star -- i.e.\ accounting for the spatial sampling -- is required. These spectra are produced as part of the pipeline, as part of the telluric correction and zero-point calibration steps.

The SAMI spectra give full coverage of the $g$ band but, due to the gap at 5700--6300\,\AA\ between the blue and red arms, only partial coverage of $r$. To complete the coverage of the $r$ band, we interpolated between the arms using a simple linear fit to the 300 pixels at the inside end of each arm. The stars selected as secondary standards, F sub-dwarfs, do not show significant features in the interpolated region.

\begin{figure}
\includegraphics[width=85mm]{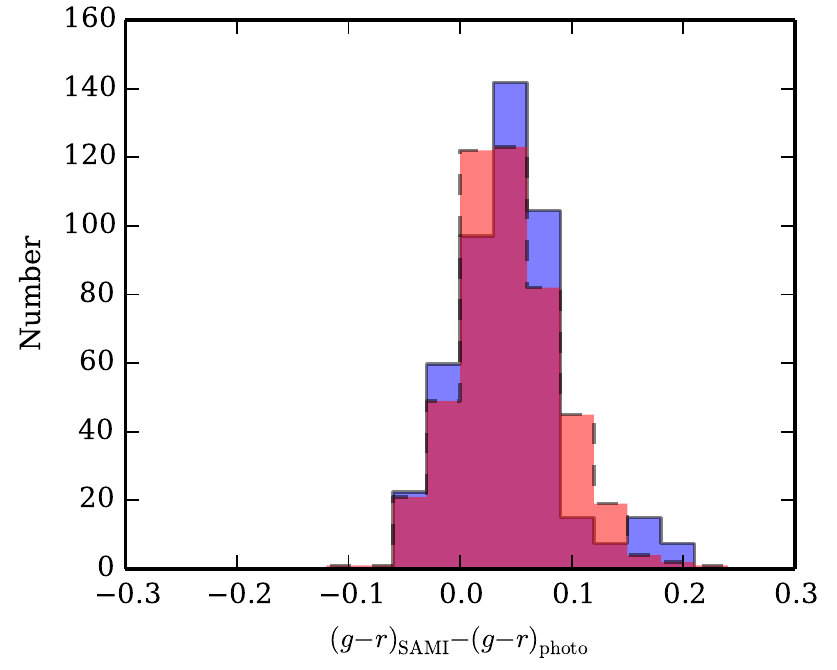}
\caption{Difference in $g-r$ colours between SAMI data and SDSS/VST-ATLAS photometry. The red histogram (dashed boundary) shows the differences for the individual RSS frames, while the blue histogram (solid boundary; scaled to have the same area as the red) shows the same for the datacubes.}
\label{fig:fluxcal_stellar_colours}
\end{figure}

After extracting the stellar spectra and interpolating between the arms, synthetic $g$ and $r$ magnitudes were derived by integrating across the SDSS filters\footnote{http://www.sdss2.org/dr7/instruments/imager/index.html\#filters} \citep{Hogg02}. The differences between the derived $g-r$ colours and those from the SDSS/VST-ATLAS catalogues, using the PSF magnitudes for SDSS and aperture magnitudes for VST-ATLAS, are shown in \mbox{Fig.\ \ref{fig:fluxcal_stellar_colours}}.

The individual RSS frames have a mean $\Delta(g-r)$ of 0.043 with standard deviation 0.040. For the final datacubes, the mean offset is 0.043 with standard deviation 0.046. In both cases the sense of the mean offset is such that the SAMI data are redder than the photometric observations. The median uncertainty in the SDSS measurements of $g-r$ colour for the stars observed is 0.024, accounting for some of the observed scatter. Converting these values to flux units indicates that the random error in the SAMI flux calibration between the $g$ and $r$ bands is only 4.3 per cent, with a systematic offset of 4.1 per cent relative to established photometry.

\subsubsection{Accuracy of zero point calibration}

\label{sec:fluxcal_zero}

Because the datacubes in each field were scaled to give the secondary standard star the same flux as measured by previous imaging surveys, the zero-point calibration of the standard stars are correct by definition. For the galaxies, a measurement of the accuracy of the calibration is possible for smaller galaxies that fully fit inside the hexabundle.

\begin{figure}
\includegraphics[width=85mm]{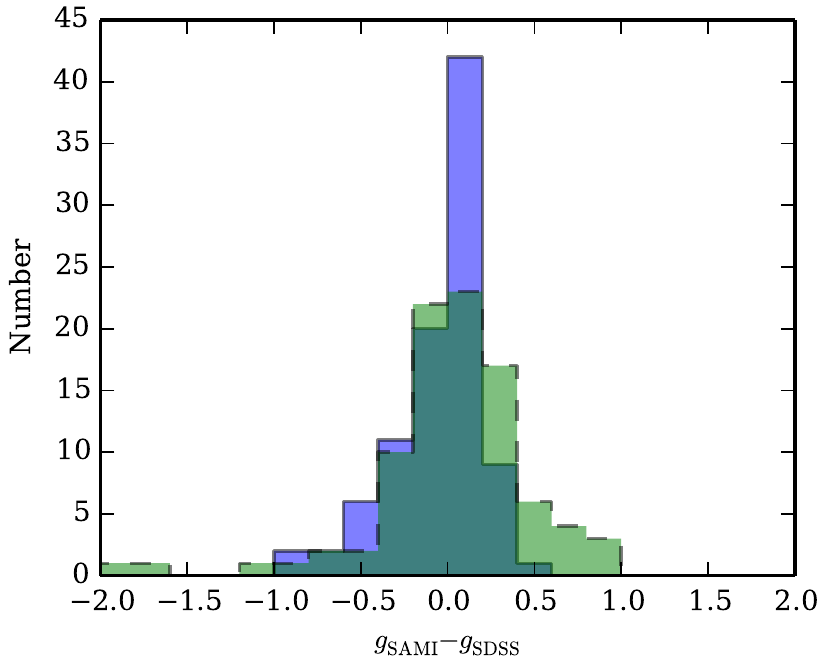}
\caption{Difference in galaxy $g$ magnitudes between SAMI data and SDSS photometry. The blue histogram (solid boundary) shows the distribution for the final datacubes, while the green histogram (dashed boundary) shows the results that are seen prior to scaling via the secondary standard star.}
\label{fig:fluxcal_galaxy_mags}
\end{figure}

\mbox{Fig.\ \ref{fig:fluxcal_galaxy_mags}} shows the distribution of offsets between the $g$-band magnitudes measured for the SAMI galaxies and those from SDSS photometry. Only the 93 galaxies in the SDSS fields with \texttt{petroR50\_r}, the radius enclosing 50 per cent of the Petrosian flux, less than 2\arcsec\ are included. Galaxies with secondary objects within their fields of view, or where the SDSS pipeline had erroneously attributed some of its flux to a star, were discarded. For the remaining galaxies, the SAMI magnitudes were found by summing all flux within the field of view, while the SDSS Petrosian magnitudes were used for comparison.

Across the sample there is a mean offset of $-$0.049\,mag ($-$4.4 per cent) and 1-$\sigma$ scatter of 0.27\,mag (28 per cent). The direction of the offset is such that the SAMI data on average are brighter than the SDSS photometry.

Also plotted in \mbox{Fig.\ \ref{fig:fluxcal_galaxy_mags}} is the distribution of $g$-band offsets prior to scaling via the secondary standard star. In this case the mean offset is 0.026\,mag (2.4 per cent) with a scatter of 0.44\,mag (50 per cent). The significant decrease in the scatter of the distribution after applying this scaling indicates that changes in atmospheric transmission affecting the entire field have a significant effect on the flux calibration, and confirms the validity of the scaling.

After scaling, the mean scatter within individual fields (including only those with at least three small galaxies) is only 0.12\,mag (12 per cent), indicating that a large part of the observed scatter is due to uncertainties in the scaling. Much of the remaining scatter may be due to the difficulty of defining magnitudes for extended, irregular sources; for illustration, we note that there is a 1-$\sigma$ scatter of 0.14\,mag between the Petrosian and model magnitudes derived by SDSS for the galaxies in the sample. Indeed, early results from a direct comparison between SDSS images and synthetic SAMI images suggest the true uncertainty is below 20 per cent, considerably smaller than the Petrosian magnitudes would indicate (Richards et al., in prep.).

\subsubsection{Telluric correction}

\label{sec:telluric}

The use of simultaneous observations of secondary standard stars allows for an accurate correction for telluric absorption in the galaxy data, with the absorption profile tailored to the atmospheric conditions at the time of observation. There are two factors that limit the accuracy of this correction: noise in the individual pixels of the extracted stellar spectrum, and systematic effects such as atmospheric variations across the 1-degree field of view.

\begin{figure}
\includegraphics[width=85mm]{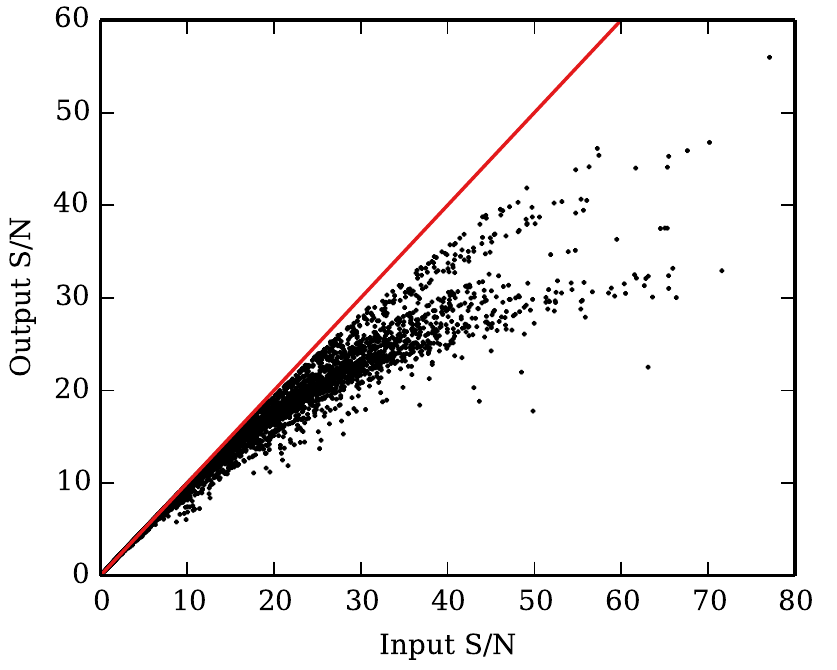}
\caption{Median output S/N within the telluric region as a function of input S/N. Each point represents a measurement from the central fibre of one galaxy in one exposure. The line of equality, representing zero loss in S/N, is plotted as a solid red line.}
\label{fig:telluric_snr}
\end{figure}

The reported per-pixel S/N of the extracted stellar spectra is typically in the range 20--50, with some observations reaching as high as 80. As the uncertainty is propagated through the telluric correction process, its effect can be quantified by the reduction in S/N of the telluric-corrected galaxy spectra within the telluric regions (6850--6960 and 7130--7360\,\AA). \mbox{Fig.\ \ref{fig:telluric_snr}} shows the median per-pixel S/N within these regions, comparing the values before (input) and after (output) telluric correction. Each point represents the central fibre from a single observation of a single galaxy. Non-central fibres follow the same pattern but typically have lower input and output S/N than central fibres. As expected, the decrease in S/N is greatest for those fibres with the highest input S/N -- a spectrum with S/N$<$10 is not degraded by a telluric correction with S/N$>$20. At higher input S/N the output S/N approaches that of the telluric correction itself. Across all central fibres, the mean decrease in S/N in the telluric bands due to the correction is 10.4 per cent.

\begin{figure}
\includegraphics[width=85mm]{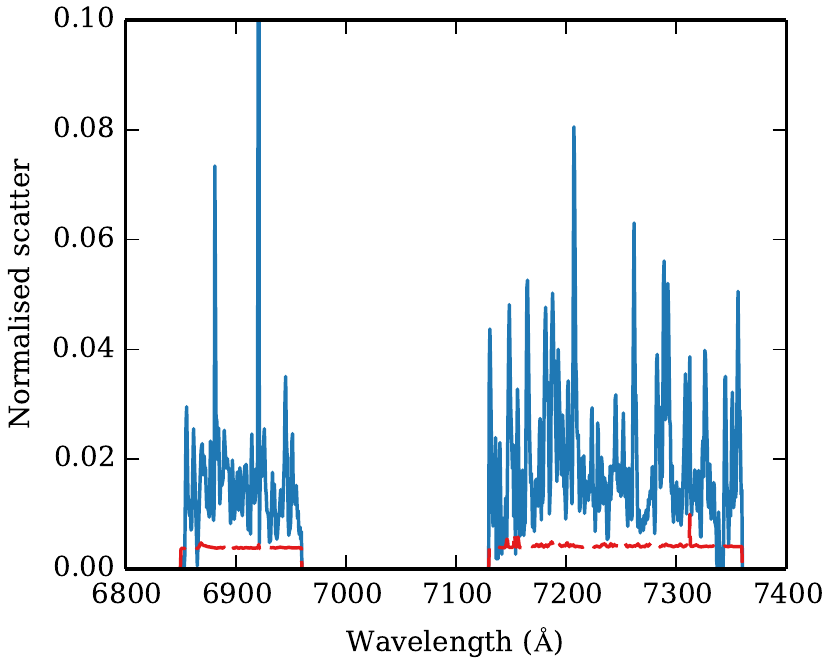}
\caption{One-$\sigma$ scatter in the measurement of the telluric transfer function, normalised to the mean transfer function, for a high-S/N observation of 13 stars. The observed scatter is plotted in blue, while the red dashed line shows the expected result if the measurement was limited by the noise in individual pixels.}
\label{fig:telluric_scatter}
\end{figure}

The level at which systematic effects limit the accuracy of the telluric correction can be measured by comparing the results derived from 13 stars observed simultaneously in a calibration field. \mbox{Fig.\ \ref{fig:telluric_scatter}} shows the 1-$\sigma$ scatter in the derived transfer function for one such field of bright stars, observed on 12 April 2013. The scatter is normalised by the mean transfer function across the 13 stars. In the same figure, the red line shows the result that would be expected from the reported S/N of the extracted spectra, i.e.\ if no systematic errors were present. We see that, although the reported S/N is $>$200 (scatter $<$0.005) at all wavelengths, the observed scatter limits the telluric correction to a S/N of 20--100 (scatter 0.01--0.05). The increase in the observed scatter is due to a combination of imperfections in the extraction process, and atmospheric variation across the 1-degree field of view. These effects mean that for higher-S/N observations the formal uncertainty may underestimate the true error within the telluric regions; however, the systematic error in the telluric correction only becomes dominant over other sources of error in the fibres with the very highest S/N.

\subsection{Datacubes}

\subsubsection{Alignment of input dataframes}

\begin{figure}
\includegraphics[width=85mm]{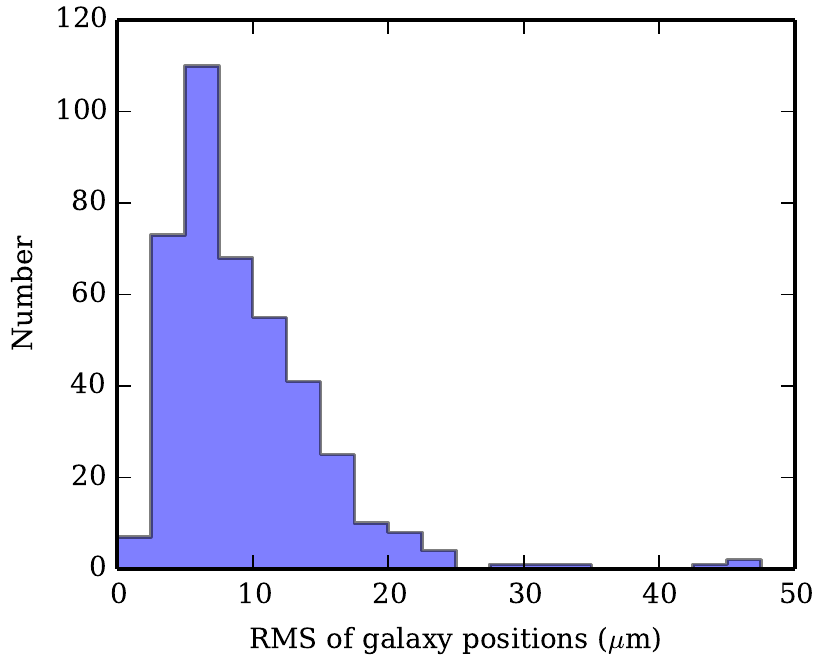}
\caption{Distribution of RMS residuals for the empirical fits to the offsets between dithered exposures.}
\label{fig:align_rms}
\end{figure}

We aligned dithered exposures by fitting a four-parameter ($x$ shift, $y$ shift, rotation and scale) empirical model to the positions of the 13 objects in each field. The accuracy of the alignment can be measured from the residuals between this model and the directly measured positions. The distribution of RMS residuals, where the RMS is taken over the 13 objects in each exposure, is shown in \mbox{Fig.\ \ref{fig:align_rms}}. By construction, the RMS cannot be above 50\,$\mu$m (0\farcs76), as in those cases the object with the most discrepant position was rejected from the fit. However, this has only been required in one frame out of the 407 observed to date. The final median RMS in the model fits, after rejecting outliers from the fit, is only 9.4\,$\mu$m (0\farcs14), less than a tenth of a fibre diameter.

\subsubsection{Point spread functions}

\label{sec:fwhm}

\begin{figure}
\includegraphics[width=85mm]{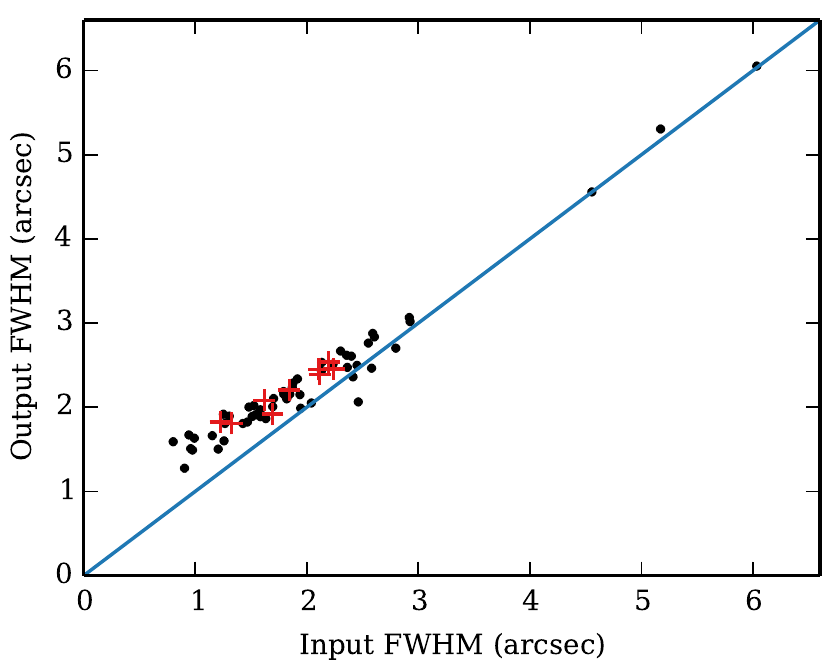}
\caption{Output vs.\ input FWHM of the PSF from Moffat function fits to the secondary standard stars. The input FWHM is a mean across fits to all RSS frames in a particular field, while the output FWHM is measured from a collapsed image of the datacube. Red crosses highlight the fields included in the EDR. The line of equality is shown in blue.}
\label{fig:fwhm_change}
\end{figure}

The spatial PSF of the SAMI datacubes is primarily determined by the atmospheric conditions at the AAT. The process of sampling an image through 1\farcs6 fibres before resampling onto a regular grid also acts to increase the final PSF. \mbox{Fig.\ \ref{fig:fwhm_change}} shows both the distribution of PSF FWHMs seen and the impact of the resampling process. The $x$ axis gives the mean FWHM of the PSF as measured from a Moffat function fit to the secondary standard star in each RSS frame, averaged over all frames for a particular field (after correcting for atmospheric dispersion). The $y$ axis shows the FWHM of the same star when measured from the finished datacube. In each case the fit accounts for the different ways in which the data are sampled.

The input FWHM measurements show a typical seeing in the range 0\farcs9--3\farcs0, with a small number of fields at higher values. For an input FWHM greater than $\sim$2\farcs5 the output FWHM follows the input very closely, but at smaller values the resampling process tends to increase the output FWHM slightly. The result is that the FWHM of the datacubes is typically in the range 1\farcs4--3\farcs0, slightly higher than the atmospheric seeing. The median FWHM in the SAMI datacubes is 2\farcs1, compared to 1\farcs8 for the input FWHM; the increase matches that expected from simulations of the resampling procedure \citep{Sharp14}. Individual cases with output FWHM less than the input value are the result of averaging over frames with large differences in input FWHM, which can occur when a field is observed over more than one night.

\subsubsection{Removal of atmospheric dispersion}

The extent of the atmospheric dispersion within each individual RSS frame was measured using the observation of the secondary standard star, as described in \citet{Sharp14}. The effect was, in principle, removed by adjusting the positions of each fibre as a function of wavelength when producing the datacube, but a residual dispersion may remain due to the uncertainties in fitting the stellar data. To measure the residual dispersion we produced two summed images of each secondary standard star from the 100 pixels centred on 4200 and 7100\,\AA, respectively. A Moffat profile was fit to each image and the distance between the centres measured.

\begin{figure}
\includegraphics[width=85mm]{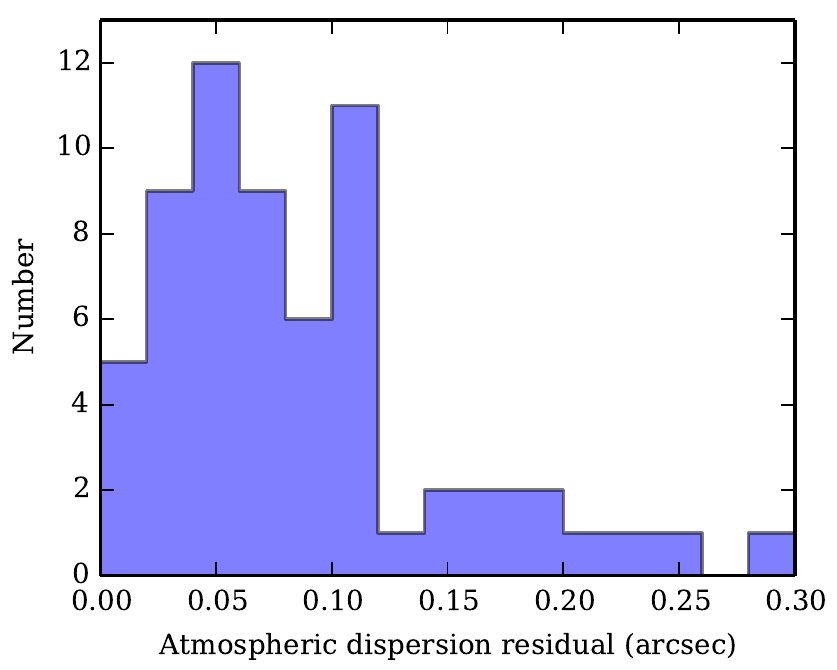}
\caption{Distribution of residual atmospheric dispersion offsets, parameterised as the angular distance between the centroids of collapsed images of the secondary standard stars at 4200 and 7100\,\AA.}
\label{fig:dar_residuals}
\end{figure}

The distribution of residual offsets is shown in \mbox{Fig.\ \ref{fig:dar_residuals}}. The mean (median) offset is only 0\farcs09 (0\farcs07), compared to an uncorrected offset of $\sim$1\arcsec\ for the typical airmass of the observations. In all cases the residual offset is less than the 0\farcs5 pixel size of the datacubes.

\subsubsection{Signal-to-noise ratio}

The spaxel size in the SAMI datacubes is a balance between the competing requirements of spatial sampling and S/N, as smaller spaxels overlap with fewer fibres and hence have lower S/N. We have chosen a relatively small spaxel size in order to conserve as much spatial information as possible, with the option to combine adjacent spaxels to increase S/N where necessary.

\begin{figure*}
\includegraphics[width=175mm]{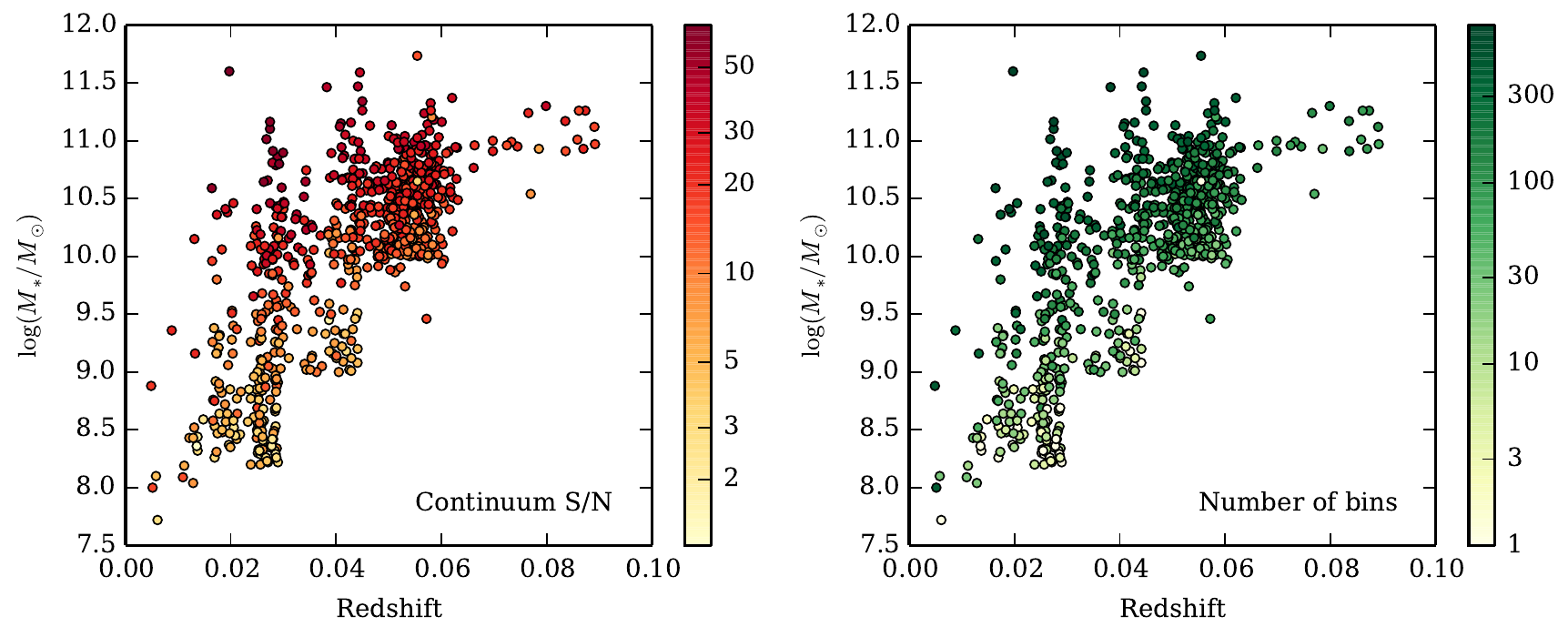}
\caption{Left: Mean continuum S/N within a radius of 1\arcsec\ centred on each galaxy, across the stellar mass vs.\ redshift plane. Right: Number of (non-independent) bins formed by the adaptive binning scheme with a continuum S/N limit of 10, across the same plane.}
\label{fig:snr}
\end{figure*}

\mbox{Fig.\ \ref{fig:snr}} shows how the S/N in a SAMI datacube varies as a function of stellar mass and redshift. In the left panel we plot the mean continuum per-pixel S/N across our 0\farcs5$\times$0\farcs5 spaxels within a circle of 1\arcsec\ radius centred on the galaxy. The S/N in each spaxel is defined as the median value across the entire wavelength range of the blue arm, where most of the stellar continuum features are found. The median S/N across these galaxies is 16.5 (10th/90th percentiles: 4.8/33.5). As expected, the S/N is a strong function of stellar mass and redshift, with high-mass, low-redshift galaxies having the highest S/N.

For IFU observations the amount of spatial information available, after binning if necessary, is often more critical than the peak S/N. The right panel of \mbox{Fig.\ \ref{fig:snr}} illustrates the level of binning that is required in order to produce spectra with a continuum S/N of at least 10. We have applied an adaptive binning scheme that combines groups of spaxels until it achieves the required S/N \citep{Cappellari03}. The bins are not in general independent of each other, due to the effects of atmospheric seeing.

The colour coding in \mbox{Fig.\ \ref{fig:snr}} shows the number of (non-independent) bins formed in this way for each galaxy, which corresponds to the level of spatial information that is retained after applying the S/N limit. The median number of bins formed in this way is 93 (10th/90th percentiles: 9/298), indicating that for the majority of SAMI galaxies a large amount of spatial information is retained even under strict S/N limits. We note that for most galaxies the number of independent spatial elements is limited by atmospheric seeing rather than S/N: for a typical PSF FWHM in the final datacubes of 2\farcs1, the maximum number of independent spatial elements is $\sim (15.0/2.1)^2=51$.

\begin{figure*}
\includegraphics[width=175mm]{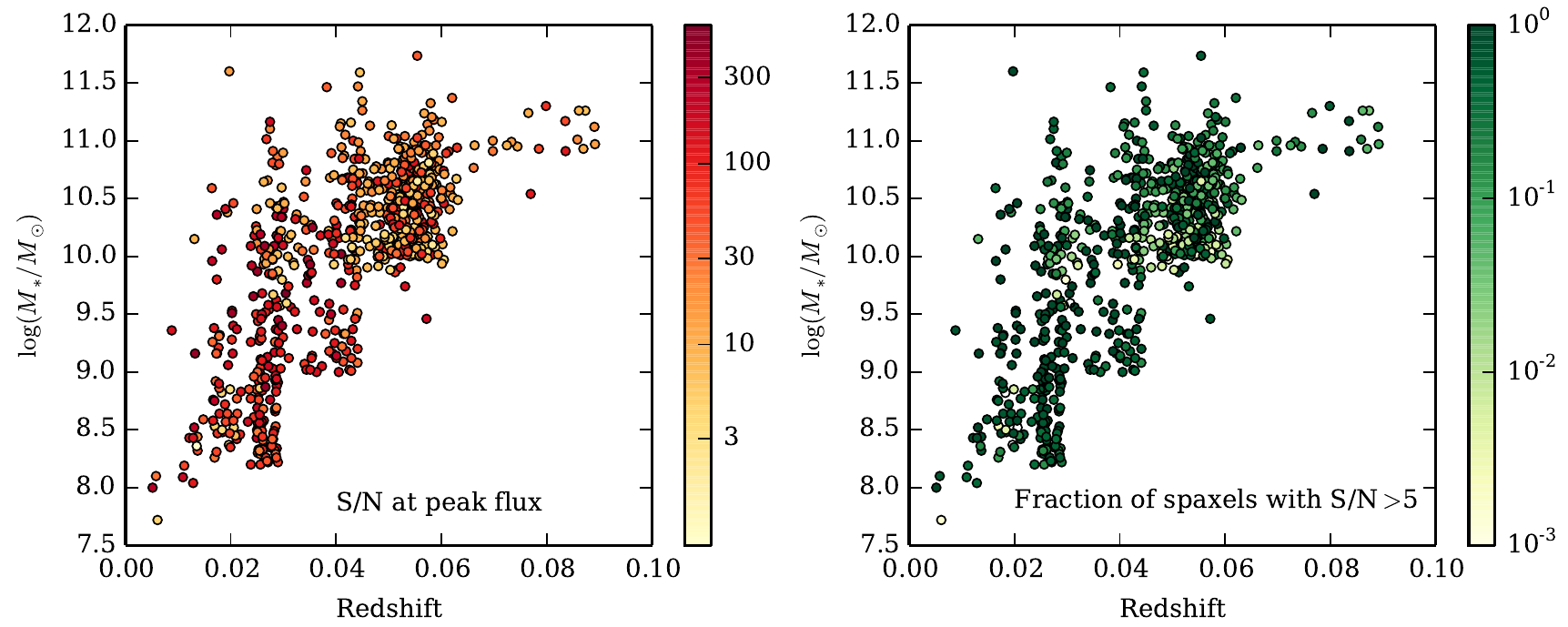}
\caption{Left: S/N of H$\alpha$ emission in the spaxel with the peak H$\alpha$ flux, across the stellar mass vs.\ redshift plane. Right: Fraction of spaxels within the field of view for which H$\alpha$ is detected with S/N$>$5.}
\label{fig:halpha_snr}
\end{figure*}

The S/N achieved in the emission lines varies enormously depending on the emission line flux in each galaxy. \mbox{Fig.\ \ref{fig:halpha_snr}} shows the per-spaxel S/N in H$\alpha$ for the peak H$\alpha$ flux (left panel), and the fraction of spaxels in the field of view for which H$\alpha$ is detected with S/N$\ge$5 (left panel). The H$\alpha$ flux and S/N were calculated from a simultaneous Gaussian fit to each strong emission line in each spectrum, after using a template fit to subtract the stellar continuum \citep{Ho14}. The median S/N measured in this way is 37.9, but the distribution ranges from $\sim$0 to over 500 (10th/90th percentiles: 5.8/156.6).

\subsubsection{Covariance}

\label{sec:covariance_effect}

\begin{figure}
\includegraphics[width=85mm]{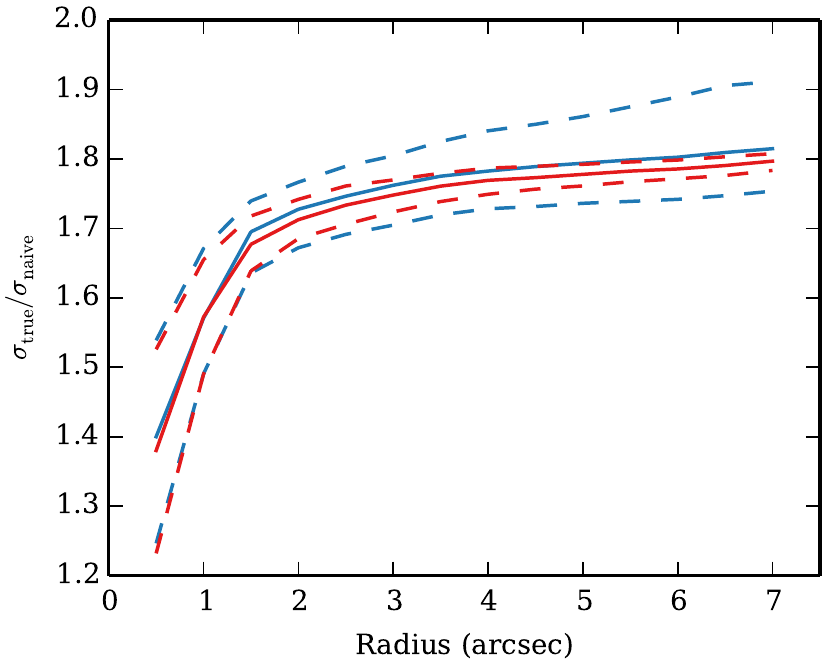}
\caption{Ratio of the true uncertainty (including covariance) to the naive uncertainty (neglecting covariance) when summing all spectra within a circular aperture. Blue and red lines give the ratios for the blue and red AAOmega arms. Solid lines show the median as a function of aperture radius for all galaxy datacubes, while the 10$^{\rm th}$ and 90$^{\rm th}$ percentiles are shown by dashed lines.}
\label{fig:covariance_effect}
\end{figure}

As described in \mbox{Section\ \ref{sec:cubing}}, the covariance between nearby spaxels must be taken into account when combining their information. Failure to do so will result in a significant underestimate of the true variance. To illustrate this point, \mbox{Fig.\ \ref{fig:covariance_effect}} shows the effect of covariance when summing spectra within a circular aperture of increasing radius. The true uncertainty (square root of variance) is calculated according to:
\begin{equation}
{\rm Var}_\lambda = \sum_{i,j}{\rm Var}_{i,j,\lambda} + 2\sum_{i,j}\sum_{k<i,l<j}{\rm Cov}_{i,j,k,l,\lambda},
\end{equation}
where Var$_\lambda$ is the variance of the summed flux as a function of wavelength, Var$_{i,j,\lambda}$ is the variance of the $(i,j)^{\rm th}$ spaxel, and Cov$_{i,j,k,l,\lambda}$ is the covariance between the $(i,j)^{\rm th}$ and $(k,l)^{\rm th}$ spaxels. The sum over covariance values runs over all unique pairs of spaxels.

In contrast, the naive uncertainty, neglecting covariance, is found simply from:
\begin{equation}
{\rm Var}_\lambda = \sum_{i,j}{\rm Var}_{i,j,\lambda}.
\end{equation}
\mbox{Fig.\ \ref{fig:covariance_effect}} shows how the ratio of the true and naive uncertainties varies as the aperture for summation, and hence number of spaxels, increases. The ratio is plotted as the median across all wavelengths within each AAOmega arm and across all galaxies in the sample. At a radius of 0\farcs5 (4 spaxels) the covariance has already increased the uncertainty by a factor of $\simeq1.4$. The ratio increases sharply to a value of $\simeq1.7$ at a radius of 1\farcs5 (32 spaxels), then slowly increases to $\simeq1.8$ as the rest of the field of view is included. The median values for the blue and red arms are very close to each other at all radii, with the blue arm having a slightly higher ratio.

\section{Conclusions}

\label{sec:conclusions}

In this paper we have described the contents of the SAMI Early Data Release (EDR). The EDR consists of a sample of 107 galaxies selected from the GAMA fields of the full SAMI Galaxy Survey sample. The galaxies in the EDR span a wide range in mass ($8.2<\log(M_*/{\rm M}_\odot)<11.6$) and redshift ($0.01<z<0.09$), and are representative of the GAMA (field and group) regions of the whole survey.

Datacubes for each galaxy covering the SAMI field of view (15\arcsec) and the wavelength ranges 3700--5600 and 6300--7400\,\AA, with a 0\farcs5 spaxel size, are available for download from the SAMI Galaxy Survey website. As well as the flux cube, the full variance and covariance information is provided. We emphasise that the covariance between nearby spaxels should be taken into account in any further analysis, as failure to do so may result in uncertainties being significantly underestimated.

A quantitative analysis of the SAMI data reduction pipeline has shown it to be performing to a very high standard at all stages. Key quality metrics include:
\begin{enumerate}
\item Cross-talk between adjacent fibres on the CCD is typically \mbox{$\sim$0.5 per cent}.
\item Flat-fielding accuracy of 0.5--1 per cent in the blue arm, rising to $\sim$2 per cent at wavelengths below 4000\,\AA, and 0.3--0.4 per cent in the red.
\item Residuals in the wavelength calibration at a level of 0.1 pixels (RMS) or better.
\item Uncertainties in the throughput calibration less than \mbox{1 per cent}.
\item After sky subtraction, median residual sky line fluxes of 0.8 per cent (blue arm) and 0.9 per cent (red arm), relative to the unsubtracted flux.
\item Continuum residuals are typically 1.2 per cent (blue arm) and 0.9 per cent (red arm) of the sky level, rising to $\sim$5 per cent in some fibres.
\item In terms of the $g-r$ colour, flux calibration that agrees with existing photometric catalogues to within 4.3 per cent, with a mean offset of 4.1 per cent (with the SAMI observations being redder).
\item An absolute flux calibration with a mean offset of 4.4 per cent (with the SAMI observations being brighter) relative to SDSS photometry, with a scatter of 28 per cent, although the true uncertainty may be considerably smaller.
\item Multiple dithered exposures for each field are aligned with a median accuracy of 9.4\,$\mu$m, less than a tenth of a fibre diameter.
\item The observed PSF in the final datacubes is primarily determined by the atmospheric conditions, with a median seeing FWHM of 2\farcs1.
\item The effects of atmospheric dispersion removed from the datacubes with a mean residual between images at different wavelengths of 0\farcs09.
\item A median per-pixel continuum S/N in the central 0\farcs5$\times$0\farcs5 spaxels of 16.5, with 10th/90th percentiles of 4.8/33.5.

\end{enumerate}

\section*{Acknowledgments}

We thank the referee for a number of useful suggestions that improved the quality of this work.

The SAMI Galaxy Survey is based on observations made at the Anglo-Australian Telescope. The Sydney-AAO Multi-object Integral field spectrograph (SAMI) was developed jointly by the University of Sydney and the Australian Astronomical Observatory. The SAMI input catalogue is based on data taken from the Sloan Digital Sky Survey, the GAMA Survey and the VST ATLAS Survey. The SAMI Galaxy Survey is funded by the Australian Research Council Centre of Excellence for All-sky Astrophysics (CAASTRO), through project number CE110001020, and other participating institutions. The SAMI Galaxy Survey website is http://sami-survey.org/ .

GAMA is a joint European-Australasian project based around a spectroscopic campaign using the Anglo-Australian Telescope. The GAMA input catalogue is based on data taken from the Sloan Digital Sky Survey and the UKIRT Infrared Deep Sky Survey. Complementary imaging of the GAMA regions is being obtained by a number of independent survey programs including GALEX MIS, VST KiDS, VISTA VIKING, WISE, Herschel-ATLAS, GMRT and ASKAP providing UV to radio coverage. GAMA is funded by the STFC (UK), the ARC (Australia), the AAO, and the participating institutions. The GAMA website is http://www.gama-survey.org/ .

JTA acknowledges the award of an Australian Research Council (ARC) Super Science Fellowship (FS110200013). SMC acknowledges the support of an ARC Future Fellowship (FT100100457). ISK is the recipient of a John Stocker Postdoctoral Fellowship from the Science and Industry Endowment Fund (Australia). MSO acknowledges the funding support from the ARC through a Super Science Fellowship (FS110200023). LC acknowledges support under the ARC Discovery Projects funding scheme (DP130100664).

This research made use of Astropy, a community-developed core Python package for Astronomy \citep{Astropy13}.


\begin{thebibliography}{}
\footnotesize{
\bibitem[\protect\citeauthoryear{{Allen} et~al.,}{{Allen}
  et~al.}{2014}]{Allen14}
{Allen} J.~T.,  et~al., 2014, Astrophysics Source Code Library, ascl:1407.006

\bibitem[\protect\citeauthoryear{{Astropy Collaboration} et~al.,}{{Astropy
  Collaboration}  et~al.}{2013}]{Astropy13}
{Astropy Collaboration} et~al., 2013, \aap, 558, A33

\bibitem[\protect\citeauthoryear{{Baldry} et~al.,}{{Baldry}
  et~al.}{2012}]{Baldry12}
{Baldry} I.~K.,  et~al., 2012, \mnras, 421, 621

\bibitem[\protect\citeauthoryear{{Bland-Hawthorn} et~al.,}{{Bland-Hawthorn}
  et~al.}{2011}]{BlandHawthorn11}
{Bland-Hawthorn} J.,  et~al., 2011, Optics Express, 19, 2649

\bibitem[\protect\citeauthoryear{{Brough} et~al.,}{{Brough}
  et~al.}{2013}]{Brough13}
{Brough} S.,  et~al., 2013, \mnras, 435, 2903

\bibitem[\protect\citeauthoryear{{Bryant}, {O'Byrne}, {Bland-Hawthorn} \&
  {Leon-Saval}}{{Bryant} et~al.}{2011}]{Bryant11}
{Bryant} J.~J.,  {O'Byrne} J.~W.,  {Bland-Hawthorn} J.,    {Leon-Saval} S.~G.,
  2011, \mnras, 415, 2173

\bibitem[\protect\citeauthoryear{{Bryant}, {Bland-Hawthorn}, {Fogarty},
  {Lawrence} \& {Croom}}{{Bryant} et~al.}{2014a}]{Bryant14a}
{Bryant} J.~J.,  {Bland-Hawthorn} J.,  {Fogarty} L.~M.~R.,  {Lawrence} J.~S.,
   {Croom} S.~M.,  2014a, \mnras, 438, 869

\bibitem[\protect\citeauthoryear{{Bryant} et~al.,}{{Bryant}
  et~al.}{2014b}]{Bryant14b}
{Bryant} J.~J.,  et~al., 2014b, \mnras\ submitted, arXiv:1407.7335

\bibitem[\protect\citeauthoryear{{Cappellari} \& {Copin}}{{Cappellari} \&
  {Copin}}{2003}]{Cappellari03}
{Cappellari} M.,  {Copin} Y.,  2003, \mnras, 342, 345

\bibitem[\protect\citeauthoryear{{Cappellari} \& {Emsellem}}{{Cappellari} \&
  {Emsellem}}{2004}]{Cappellari04}
{Cappellari} M.,  {Emsellem} E.,  2004, \pasp, 116, 138

\bibitem[\protect\citeauthoryear{{Cappellari} et~al.,}{{Cappellari}
  et~al.}{2011a}]{Cappellari11a}
{Cappellari} M.,  et~al., 2011a, \mnras, 413, 813

\bibitem[\protect\citeauthoryear{{Cappellari} et~al.,}{{Cappellari}
  et~al.}{2011b}]{Cappellari11b}
{Cappellari} M.,  et~al., 2011b, \mnras, 416, 1680

\bibitem[\protect\citeauthoryear{{Childress}, {Vogt}, {Nielsen} \&
  {Sharp}}{{Childress} et~al.}{2014}]{Childress14}
{Childress} M.~J.,  {Vogt} F.~P.~A.,  {Nielsen} J.,    {Sharp} R.~G.,  2014,
  \apss, 349, 617

\bibitem[\protect\citeauthoryear{{Colless} et~al.,}{{Colless}
  et~al.}{2001}]{Colless01}
{Colless} M.,  et~al., 2001, \mnras, 328, 1039

\bibitem[\protect\citeauthoryear{{Croom} et~al.,}{{Croom}
  et~al.}{2012}]{Croom12}
{Croom} S.~M.,  et~al., 2012, \mnras, 421, 872

\bibitem[\protect\citeauthoryear{{Driver} et~al.,}{{Driver}
  et~al.}{2009}]{Driver09}
{Driver} S.~P.,  et~al., 2009, Astronomy and Geophysics, 50, 12

\bibitem[\protect\citeauthoryear{{Driver} et~al.,}{{Driver}
  et~al.}{2011}]{Driver11}
{Driver} S.~P.,  et~al., 2011, \mnras, 413, 971

\bibitem[\protect\citeauthoryear{{Flores}, {Hammer}, {Puech}, {Amram} \&
  {Balkowski}}{{Flores} et~al.}{2006}]{Flores06}
{Flores} H.,  {Hammer} F.,  {Puech} M.,  {Amram} P.,    {Balkowski} C.,  2006,
  \aap, 455, 107

\bibitem[\protect\citeauthoryear{{Fogarty} et~al.,}{{Fogarty}
  et~al.}{2012}]{Fogarty12}
{Fogarty} L.~M.~R.,  et~al., 2012, \apj, 761, 169

\bibitem[\protect\citeauthoryear{{Fogarty} et~al.,}{{Fogarty}
  et~al.}{2014}]{Fogarty14}
{Fogarty} L.~M.~R.,  et~al., 2014, \mnras, 443, 485

\bibitem[\protect\citeauthoryear{{Fruchter} \& {Hook}}{{Fruchter} \&
  {Hook}}{2002}]{Fruchter02}
{Fruchter} A.~S.,  {Hook} R.~N.,  2002, \pasp, 114, 144

\bibitem[\protect\citeauthoryear{{Genzel} et~al.,}{{Genzel}
  et~al.}{2008}]{Genzel08}
{Genzel} R.,  et~al., 2008, \apj, 687, 59

\bibitem[\protect\citeauthoryear{{Hill} et~al.,}{{Hill}  et~al.}{2011}]{Hill11}
{Hill} D.~T.,  et~al., 2011, \mnras, 412, 765

\bibitem[\protect\citeauthoryear{{Ho} et~al.,}{{Ho}  et~al.}{2014}]{Ho14}
{Ho} I.-T.,  et~al., 2014, \mnras, 444, 3894

\bibitem[\protect\citeauthoryear{{Hogg}, {Baldry}, {Blanton} \&
  {Eisenstein}}{{Hogg} et~al.}{2002}]{Hogg02}
{Hogg} D.~W.,  {Baldry} I.~K.,  {Blanton} M.~R.,    {Eisenstein} D.~J.,  2002,
  ArXiv e-prints, astro-ph/0210394

\bibitem[\protect\citeauthoryear{{Hopkins} et~al.,}{{Hopkins}
  et~al.}{2003}]{Hopkins03}
{Hopkins} A.~M.,  et~al., 2003, \apj, 599, 971

\bibitem[\protect\citeauthoryear{{Horne}}{{Horne}}{1986}]{Horne86}
{Horne} K.,  1986, \pasp, 98, 609

\bibitem[\protect\citeauthoryear{{Husemann} et~al.,}{{Husemann}
  et~al.}{2013}]{husemann13}
{Husemann} B.,  et~al., 2013, \aap, 549, A87

\bibitem[\protect\citeauthoryear{{Jones} et~al.,}{{Jones}
  et~al.}{2009}]{Jones09}
{Jones} D.~H.,  et~al., 2009, \mnras, 399, 683

\bibitem[\protect\citeauthoryear{{Kelvin} et~al.,}{{Kelvin}
  et~al.}{2012}]{Kelvin12}
{Kelvin} L.~S.,  et~al., 2012, \mnras, 421, 1007

\bibitem[\protect\citeauthoryear{{Konstantopoulos}}{{Konstantopoulos}}{2014}]{Konstantopoulos14}
{Konstantopoulos} I.~S.,  2014, Astronomy \& Computing submitted, arXiv:1407.5619

\bibitem[\protect\citeauthoryear{{Kron}}{{Kron}}{1980}]{Kron80}
{Kron} R.~G.,  1980, \apjs, 43, 305

\bibitem[\protect\citeauthoryear{{Lara-L{\'o}pez} et~al.,}{{Lara-L{\'o}pez} et~al.}{2013}]{LaraLopez13}
{Lara-L{\'o}pez} M.~A., et~al., 2013, \mnras, 434, 451

\bibitem[\protect\citeauthoryear{{Lewis} et~al.,}{{Lewis}
  et~al.}{2002}]{Lewis02}
{Lewis} I.,  et~al., 2002, \mnras, 334, 673

\bibitem[\protect\citeauthoryear{{Mannucci} et~al.,}{{Mannucci} et~al.}{2010}]{Mannucci10}
{Mannucci} F., {Cresci} G., {Maiolino} R., {Marconi} A., {Gnerucci} A., 2010, \mnras, 408, 2115

\bibitem[\protect\citeauthoryear{{Pasquini} et~al.,}{{Pasquini}
  et~al.}{2002}]{Pasquini02}
{Pasquini} L.,  et~al., 2002, The Messenger, 110, 1

\bibitem[\protect\citeauthoryear{{Pence}, {Chiappetti}, {Page}, {Shaw} \&
  {Stobie}}{{Pence} et~al.}{2010}]{Pence10}
{Pence} W.~D.,  {Chiappetti} L.,  {Page} C.~G.,  {Shaw} R.~A.,    {Stobie} E.,
  2010, \aap, 524, A42

\bibitem[\protect\citeauthoryear{{Petrosian}}{{Petrosian}}{1976}]{Petrosian76}
{Petrosian} V.,  1976, \apjl, 209, L1

\bibitem[\protect\citeauthoryear{{Richards} et~al.,}{{Richards} et~al.}{2014}]{Richards14}
{Richards} S.~N., et~al., 2014, \mnras\ accepted, arXiv:1409.4495

\bibitem[\protect\citeauthoryear{{S{\'a}nchez} et~al.,}{{S{\'a}nchez}
  et~al.}{2012}]{Sanchez12}
{S{\'a}nchez} S.~F.,  et~al., 2012, \aap, 538, A8

\bibitem[\protect\citeauthoryear{{S{\'a}nchez-Bl{\'a}zquez} et~al.,}
 {{S{\'a}nchez-Bl{\'a}zquez} et~al.}{2014}]{SanchezBlazquez14}
{S{\'a}nchez-Bl{\'a}zquez} P., et~al., 2014, \aap\ accepted, arXiv:1407.0002

\bibitem[\protect\citeauthoryear{{Shanks} et~al.,}{{Shanks}
  et~al.}{2013}]{Shanks13}
{Shanks} T.,  et~al., 2013, The Messenger, 154, 38

\bibitem[\protect\citeauthoryear{{Sharp} \& {Birchall}}{{Sharp} \&
  {Birchall}}{2010}]{Sharp10a}
{Sharp} R.,  {Birchall} M.~N.,  2010, \pasa, 27, 91

\bibitem[\protect\citeauthoryear{{Sharp} \& {Bland-Hawthorn}}{{Sharp} \&
  {Bland-Hawthorn}}{2010}]{Sharp10b}
{Sharp} R.~G.,  {Bland-Hawthorn} J.,  2010, \apj, 711, 818

\bibitem[\protect\citeauthoryear{{Sharp} \& {Parkinson}}{{Sharp} \&
  {Parkinson}}{2010}]{Sharp10c}
{Sharp} R.,  {Parkinson} H.,  2010, \mnras, 408, 2495

\bibitem[\protect\citeauthoryear{{Sharp} et~al.,}{{Sharp}
  et~al.}{2006}]{Sharp06}
{Sharp} R.,  et~al., 2006, in Society of Photo-Optical Instrumentation
  Engineers (SPIE) Conference Series Vol.~6269

\bibitem[\protect\citeauthoryear{{Sharp}, {Brough} \& {Cannon}}{{Sharp}
  et~al.}{2013}]{Sharp13}
{Sharp} R.,  {Brough} S.,    {Cannon} R.~D.,  2013, \mnras, 428, 447

\bibitem[\protect\citeauthoryear{{Sharp} et~al.,}{{Sharp}
  et~al.}{2014}]{Sharp14}
{Sharp} R.,  et~al., 2014, \mnras\ submitted, arXiv:1407.5237

\bibitem[\protect\citeauthoryear{{Shen}, {Vanden Berk}, {Schneider} \&
  {Hall}}{{Shen} et~al.}{2008}]{Shen08}
{Shen} J.,  {Vanden Berk} D.~E.,  {Schneider} D.~P.,    {Hall} P.~B.,  2008,
  \aj, 135, 928

\bibitem[\protect\citeauthoryear{{Taylor} et~al.,}{{Taylor}
  et~al.}{2011}]{Taylor11}
{Taylor} E.~N.,  et~al., 2011, \mnras, 418, 1587

\bibitem[\protect\citeauthoryear{{Tonry}, {Blakeslee}, {Ajhar} \&
  {Dressler}}{{Tonry} et~al.}{2000}]{Tonry00}
{Tonry} J.~L.,  {Blakeslee} J.~P.,  {Ajhar} E.~A.,    {Dressler} A.,  2000,
  \apj, 530, 625

\bibitem[\protect\citeauthoryear{{Wijesinghe} et~al.,}{{Wijesinghe}
  et~al.}{2012}]{Wijesinghe12}
{Wijesinghe} D.~B.,  et~al., 2012, \mnras, 423, 3679

\bibitem[\protect\citeauthoryear{{Yang} et~al.,}{{Yang}  et~al.}{2008}]{Yang08}
{Yang} Y.,  et~al., 2008, \aap, 477, 789

\bibitem[\protect\citeauthoryear{{York} et~al.,}{{York}  et~al.}{2000}]{York00}
{York} D.~G.,  et~al., 2000, \aj, 120, 1579}

\end{thebibliography}

\appendix

\section{Galaxies in the SAMI EDR}

Table \ref{tab:galaxies} gives information about the galaxies included in the SAMI Galaxy Survey EDR.

\begin{table*}
\caption{Galaxies included in the SAMI EDR. For each galaxy we list: the galaxy name; J2000 coordinates (decimal degrees); extinction-corrected $r$-band magnitudes measured using the Petrosian (`petro'; \citealt{Petrosian76}) and Kron (`auto'; \citealt{Kron80}) systems \citep{Hill11}; redshifts corrected (`tonry'; \citealt{Tonry00,Baldry12}) and uncorrected (`spec') for large-scale flow; $r$-band rest-frame absolute magnitude; $r$-band effective radius ($R_e$) along major axis (arcsec; \citealt{Kelvin12}); $r$-band surface brightness (mag\,arcsec$^{-2}$) within $R_e$, at $R_e$ and at $2R_e$; ellipticity and position angle (deg), measured in the $r$ band; logarithm of stellar mass (M$_\odot$; \citealt{Taylor11,Bryant14b}); $g-i$ colour from Kron magnitudes; Galactic extinction in the $g$ band; GAMA catalogue ID; SAMI Galaxy Survey priority (8: main sample, 4: high-mass fillers, 3: other fillers); field in which the galaxy was observed; and URLs for the blue and red datacubes. Only the first five entries are printed; the full table is available as Supporting Information with the online version of the paper, or on the SAMI Galaxy Survey website at http://sami-survey.org/edr/data/SAMI\_EarlyDataRelease.txt .}
\label{tab:galaxies}
\begin{tabular}{rrrrrrrrr}
\hline
\multicolumn{1}{c}{Name} & \multicolumn{1}{c}{R.A.\ (deg)} & \multicolumn{1}{c}{Dec.\ (deg)} & \multicolumn{1}{c}{$r_{\rm petro}$} & \multicolumn{1}{c}{$r_{\rm auto}$} & \multicolumn{1}{c}{$z_{\rm tonry}$} & \multicolumn{1}{c}{$z_{\rm spec}$} & \multicolumn{1}{c}{$M_{\rm r}$} & \multicolumn{1}{c}{$R_{\rm e}$ (arcsec)} \\
\hline
J084445.41$+$021107.8 & 131.1892 & $+$2.1855 & 15.781 & 15.801 & 0.07548 & 0.07444 & $-$21.96 & 4.17 \\
J084458.94$+$020305.9 & 131.2456 & $+$2.0517 & 16.167 & 16.153 & 0.05193 & 0.05091 & $-$20.70 & 1.94 \\
J084459.22$+$022834.8 & 131.2468 & $+$2.4763 & 15.981 & 15.942 & 0.05064 & 0.04962 & $-$20.88 & 6.76 \\
J084517.68$+$021650.0 & 131.3237 & $+$2.2806 & 17.036 & 17.021 & 0.05106 & 0.05004 & $-$19.81 & 2.87 \\
J084531.79$+$022833.1 & 131.3825 & $+$2.4759 & 15.311 & 15.188 & 0.07756 & 0.07651 & $-$22.64 & 6.76 \\
\hline
\end{tabular}
\end{table*}
\begin{table*}
\contcaption{}
\begin{tabular}{rrrrr}
\hline
\multicolumn{1}{c}{$\langle\mu(R_{\rm e})\rangle$ (mag\,arcsec$^{-2}$)} & \multicolumn{1}{c}{$\mu(R_{\rm e})$ (mag\,arcsec$^{-2}$)} & \multicolumn{1}{c}{$\mu(2R_{\rm e})$ (mag\,arcsec$^{-2}$)} & \multicolumn{1}{c}{Ellip.} & \multicolumn{1}{c}{P.A.\ (deg)} \\
\hline
20.16 & 21.40 & 23.00 & 0.5116 & 3.28 \\
19.41 & 20.54 & 22.17 & 0.2758 & 93.58 \\
21.02 & 22.49 & 24.06 & 0.5711 & 8.36 \\
21.31 & 22.58 & 24.18 & 0.0120 & 11.30 \\
21.10 & 22.43 & 24.01 & 0.2140 & 160.43 \\
\hline
\end{tabular}
\end{table*}
\begin{table*}
\contcaption{}
\begin{tabular}{rrrrcc}
\hline
\multicolumn{1}{c}{log($M_*/{\rm M}_\odot$)} & \multicolumn{1}{c}{$g-i$} & \multicolumn{1}{c}{$A_{\rm g}$} & \multicolumn{1}{c}{CATAID} & \multicolumn{1}{c}{SURV\_SAMI} & \multicolumn{1}{c}{Tile no.} \\
\hline
10.95 & 1.30 & 0.244 & 386268 & 8 & Y13SAR1\_P005\_09T009 \\
10.26 & 1.00 & 0.316 & 345820 & 8 & Y13SAR1\_P005\_09T009 \\
10.36 & 1.02 & 0.154 & 517164 & 8 & Y13SAR1\_P005\_09T009 \\
10.09 & 1.22 & 0.212 & 417486 & 8 & Y13SAR1\_P005\_09T009 \\
11.24 & 1.31 & 0.165 & 517205 & 8 & Y13SAR1\_P005\_09T009 \\
\hline
\end{tabular}
\end{table*}
\begin{table*}
\contcaption{}
\begin{tabular}{l}
\hline
\multicolumn{1}{c}{Blue cube} \\
\hline
http://sami-survey.org/edr/data/386268/386268\_blue\_7\_Y13SAR1\_P005\_09T009.fits.gz \\
http://sami-survey.org/edr/data/345820/345820\_blue\_7\_Y13SAR1\_P005\_09T009.fits.gz \\
http://sami-survey.org/edr/data/517164/517164\_blue\_7\_Y13SAR1\_P005\_09T009.fits.gz \\
http://sami-survey.org/edr/data/417486/417486\_blue\_7\_Y13SAR1\_P005\_09T009.fits.gz \\
http://sami-survey.org/edr/data/517205/517205\_blue\_7\_Y13SAR1\_P005\_09T009.fits.gz \\
\hline
\end{tabular}
\end{table*}
\begin{table*}
\contcaption{}
\begin{tabular}{l}
\hline
\multicolumn{1}{c}{Red cube} \\
\hline
http://sami-survey.org/edr/data/386268/386268\_red\_7\_Y13SAR1\_P005\_09T009.fits.gz \\
http://sami-survey.org/edr/data/345820/345820\_red\_7\_Y13SAR1\_P005\_09T009.fits.gz \\
http://sami-survey.org/edr/data/517164/517164\_red\_7\_Y13SAR1\_P005\_09T009.fits.gz \\
http://sami-survey.org/edr/data/417486/417486\_red\_7\_Y13SAR1\_P005\_09T009.fits.gz \\
http://sami-survey.org/edr/data/517205/517205\_red\_7\_Y13SAR1\_P005\_09T009.fits.gz \\
\hline
\end{tabular}
\end{table*}

\end{document}